\documentclass[fleqn,usenatbib]{mnras}

\usepackage{newtxtext,newtxmath}

\usepackage[T1]{fontenc}

\DeclareRobustCommand{\VAN}[3]{#2}
\let\VANthebibliography\thebibliography
\def\thebibliography{\DeclareRobustCommand{\VAN}[3]{##3}\VANthebibliography}

\usepackage{graphicx}	
\usepackage{amsmath}	
\usepackage{multirow}
\usepackage{pdflscape}
\usepackage{ulem}
\usepackage{bm}

\title[Shallower radius valley around low-mass hosts]{Shallower radius valley around low-mass hosts: Evidence for icy planets, collisions or high-energy radiation scatter}

\author[C. S. K. Ho et al.]{
Cynthia S. K. Ho$^{1,2}$\thanks{E-mail: sze.ho.20@ucl.ac.uk},
James G. Rogers$^{3}$,
Vincent Van Eylen$^{1,2}$,
James E. Owen$^{4}$
and Hilke E. Schlichting$^{3}$
\\
$^{1}$Mullard Space Science Laboratory, University College London, Dorking, RH5 6NT, UK \\
$^{2}$The Centre for Planetary Sciences at UCL/Birkbeck, London WC1E 6BT, UK \\
$^{3}$Department of Earth, Planetary, and Space Sciences, The University of California, Los Angeles, 595 Charles E. Young Drive East, Los Angeles, CA 90095, USA \\
$^{4}$Imperial Astrophysics, Department of Physics, Imperial College London, Prince Consort Rd, London, SW7 2AZ, UK }

\date{Accepted 2024 May 24. Received 2024 May 24; in original form 2023 December 2018}

\pubyear{2023}

\begin{document}
\label{firstpage}
\pagerange{\pageref{firstpage}--\pageref{lastpage}}
\maketitle

\begin{abstract}
The radius valley, i.e., a dearth of planets with radii between 1.5 and 2 Earth radii, provides insights into planetary formation and evolution. Using homogenously revised planetary parameters from Kepler 1-minute short cadence light curves, we remodel transits of 72 small planets mostly orbiting low-mass stars, improving the precision and accuracy of planet parameters. By combining this sample with a similar sample of planets around higher-mass stars, we determine the depth of the radius valley as a function of stellar mass. We find that the radius valley is shallower for low-mass stars compared to their higher mass counterparts. Upon comparison, we find that theoretical models of photoevaporation under-predict the number of planets observed inside the radius valley for low-mass stars: with decreasing stellar mass, the predicted fraction of planets inside the valley remains approximately constant whereas the observed fraction increases. We argue that this provides evidence for the presence of icy planets around low-mass stars. Alternatively, planets orbiting low-mass stars undergo more frequent collisions and scatter in the stars' high-energy output may also cause planets to fill the valley. We predict that more precise mass measurements for planets orbiting low mass stars would be able to distinguish between these scenarios.
\end{abstract}



\defcitealias{ho2023deep}{H23}

\section{Introduction}
The `radius valley' is an observed paucity of planets between about 1.5 and 2 Earth radii. Its properties provide insights into planetary formation and evolution mechanisms. One group of theories suggests that small, close-in planets undergo thermally-driven mass loss scenarios, including photoevaporation \citep[e.g.][]{owen2013kepler, lopez2013role, chen2016evolutionary, owen2017evaporation}, and core-powered mass loss \citep[e.g.][]{ginzburg2018core, gupta2019sculpting, gupta2020signatures}, resulting in a separation between planets without substantial atmospheres, and planets that have retained them. Alternatively, it is suggested that under the late gas-poor formation scenario, planets below the valley have formed after most of the gas has dissipated and hence do not have an atmosphere \citep[e.g.][]{lopez2018how}.

The radius valley around FGK stars has been extensively observed \citep[e.g.][]{fulton2017california,  vaneylen2018asteroseismic, fulton2018california}. These observations all show a separation between planets above the valley, which are thought to possess a substantial hydrogen-helium (H-He) atmosphere, and planets below the valley, which are expected to be bare rocky cores. These studies have also found that the radius valley shifts to a larger radius for shorter orbital periods, which is consistent with thermally-driven mass loss models, as larger planets could lose their atmospheres closer to their hosts. The radius valley is also found to shift to a larger radius for more massive host stars, which again agrees with both photoevaporation and core-powered mass loss models \citep[e.g.][]{berger2020gaia2, petigura2022california, ho2023deep}.

However, the picture is different when it comes to M-dwarf host stars. \citet{cloutier2020evolution} studied the radius valley around low-mass stars, and found that the valley location tends to larger planet radii for larger planet-star separation, opposite of the dependence for FGK stars. They suggested that this is evidence for such planets to undergo gas-poor formation \citep[e.g.][]{lopez2018how, cloutier2020evolution}. Yet, \citet{vaneylen2021masses} analysed small planets orbiting M-dwarf stars and found the valley location tending to smaller planet radii for larger planet-star separation, similar to that of FGK stars, contradicting the findings of \citet{cloutier2020evolution}. \citet{bonfanti2024characterising} also similarly analysed the orbital period dependence for the M-dwarf radius valley, and found a similar, but much weaker dependence, than \citet{vaneylen2021masses}. Recently, \citet{luque2022density} investigated the distribution of 43 small transiting planets around M dwarfs with mass measurements and argued that there is no orbital period dependence of the radius valley, but rather planets can be clearly separated into three distinct groups in bulk density space: rocky planets, `water worlds' with 50\% water, and `puffy' sub-Neptunes with sizeable atmospheres. Similarly, studies from \citet{venturini2020nature,mousis2020irradiated,venturini2024fading} have suggested that the valley emerges due to a dichotomy in density between supercritical water worlds and rocky planets. However, \citet{rogers2023conclusive} counter-argued that thermally-driven mass loss mechanisms can still explain the distribution of small planets around low-mass stars. If a `boil-off' scenario existed, i.e. a significant amount of H-He atmosphere is removed during disk dispersal, the resulting planet distribution matches well with that of the `water worlds' identified in \citet{luque2022density}. Even if the `boil-off' process does not happen, and the initial atmospheric loss processes are not known, the resulting distribution is consistent with the overall arrangement noted in \citet{luque2022density}. This is because the atmospheric mass fraction scales with the planet mass, when atmospheric mass-loss is taken into account \citep{rogers2023conclusive}. Hence, it is uncertain whether the difference in the valley's dependence with orbital period can be explained by the existence of such `water worlds'. These discrepancies in the characteristics of the radius valley for host stars of different masses, as well as the various views and interpretations of the radius valley for low-mass stars, suggest the possibility that stellar mass plays a role in determining the evolutionary pathways of planets. 

\citet{ho2023deep} found that careful fitting of short-cadence transit light curves to obtain precise planetary radii caused some inferred planet properties to change \citep[see also][]{camero2023understanding}, revealing a deeper, more sharply-determined radius valley. \cite{ho2023deep} showed that the radius valley properties depend on stellar mass, but focused on planets orbiting FGK stars. In this work, we extend the sample by conducting a similar, careful and homogeneous analysis of \textit{Kepler} short cadence transit observations to study the radius valley for planets orbiting low-mass M-type stars. 
We compare the observed radius valley for stars of different masses and attempt to reproduce the observed radius valley characteristics using photoevaporation models.

In Section~\ref{sect:method}, we describe the planet sample and the methods used to fit the \textit{Kepler} transit light curves to obtain new planetary parameters. In Section~\ref{sect:results}, we analyse the observed radius valley's change with stellar mass and compare them with theoretical models. The physical implications of the results are discussed in Section~\ref{sect:discussion}. Finally, we provide conclusions and outlook in Section~\ref{sect:conclusions}.

\section{Properties of planets orbiting around low-mass stars} \label{sect:method}
\subsection{Sample of new planets orbiting low-mass stars} \label{subsect:sample_sel}
In this work, we study planets around low-mass stars, using planets in the sample from the California-\textit{Kepler} Survey (CKS) `extended mass' (CXM) sample \citep{petigura2022california}, which has a stellar mass range between $0.4$ and $1.51M_{\sun}$. The CXM is a filtered sample of the union of the Kepler Data Release 25 stellar properties catalogue \citep{mathur2017revised}, the Gaia DR2 catalogue \citep{gaia2018gaia}, and the Gaia-Kepler Stellar Properties Catalogue \citep{berger2020gaia1}. 

We select planets that meet the criteria defined in \citet{ho2023deep}, i.e. 
\begin{enumerate}
    \item $1 \leq R_{\text{p}}/R_{\earth} \leq 4$, where $R_{\text{p}}$ is the planetary radius,
    \item $1 \leq P/\text{d} \leq 100$, where $P$ is the orbital period,
    \item at least 6 months of \textit{Kepler} 1-minute short cadence data available, and
    \item without transit timing variations (TTV) measurements, as reported in \citet{holczer2016transit}.
\end{enumerate}
This gives us 72 planets not previously refitted by \cite{ho2023deep} for which we homogeneously refit the planetary transits to update the planetary parameters. While the same orbital period does not correspond to the same irradiation received by the planet around stars of different masses, the radius valley is orbital-period dependent \citep[e.g.][]{vaneylen2018asteroseismic,ho2023deep}, hence we expect that our analysis, which accounts for this dependency, remains robust across various stellar mass ranges. In our subsequent analysis, we will divide our sample into multiple stellar mass bins, therefore minimising the difference in irradiation received by the planet for the same orbital period. Furthermore, the radius valley lies well within $1 \leq R_{\text{p}}/R_{\earth} \leq 4$ for the stellar mass ranges studied in this work \citep[e.g.][]{ho2023deep}, therefore we do not expect the radius cuts implemented during sample filtering to significantly impact our analysis of the radius valley.

\subsection{Transit Fitting} 
Since \textit{Kepler} 1-minute short cadence light curves produce superior planetary radius precision due to their higher sensitivity towards changes in the transit shape \citep[e.g.][]{petigura2020two, ho2023deep, camero2023understanding}, we refit the 72 planets that have not been studied in \citet{ho2023deep} using \textit{Kepler} 1-minute cadence light curves, to obtain updated planetary parameters. 

We follow the transit fitting method used in \citet{ho2023deep} and summarise its main characteristics here. We start by performing data reduction and light curve detrending, then correcting the light curves with the `radius correction factor' \citep{furlan2017kepler} to account for stellar multiplicities. We then use \texttt{exoplanet} \citep{foremanmackey2021exoplanet} to generate a transit light curve model with quadratic stellar limb darkening, and run a Hamiltonian Monte Carlo (HMC) algorithm implemented in \texttt{PyMC3} \citep{salvatier2016probabilistic} with a Gaussian Process (GP) model \citep{rasmussen2006gaussian} included to account for undetected noise in the light curves. We initialise the HMC chains in the same way and apply the same priors as in \citet{ho2023deep}. Similarly, we fit for the following parameters: orbital period ($P$), transit mid-time ($t_0$), planet-to-star radius ratio ($R_{\text{p}}/R_{\star}$), impact parameter ($b$), eccentricity ($e$), argument of periapsis ($\omega$), stellar density ($\rho_{\star}$), and two quadratic stellar limb darkening parameters ($u_0$ and $u_1$). We use the values from \citet{petigura2022california} as the initial guesses for $\rho_{\star}$, as these stars are not included in the \citet{fulton2018california} sample. The initial guesses for $u_0$ and $u_1$ are estimated using the Exoplanet Characterization ToolKit \citep{bourque2021exoplanet}, with the input effective temperature ($T_{\text{eff}}$), surface gravity ($\log{g}$), and metallicity ([Fe/H]) taken from the \textit{Kepler} Q1-16 dataset \citep{mullally2015planetary}.

\subsection{Revised planetary parameters} \label{subsect:results_params}
We report the revised planetary parameters of all 72 planets in Table~\ref{tab:planet_params}.  For subsequent analysis of the radius valley, we only consider those planets whose properties are well-known. Following \citet{ho2023deep}, we, therefore, apply the following parameter constraints:
\begin{enumerate}
    \item measurement uncertainty on the planet radius ($\sigma_{R_\text{p}}) \leq 10\%$,
    \item radius correction factor (RCF) $\leq 5\%$, as reported in \citet{furlan2017kepler}; this is to account for stellar flux from nearby stars contaminating the light curves \citep{furlan2017kepler}, and
    \item at least 3 transits in the \textit{Kepler} short cadence data.
\end{enumerate}

\begin{figure}
    \centering
    \includegraphics[width=\columnwidth]{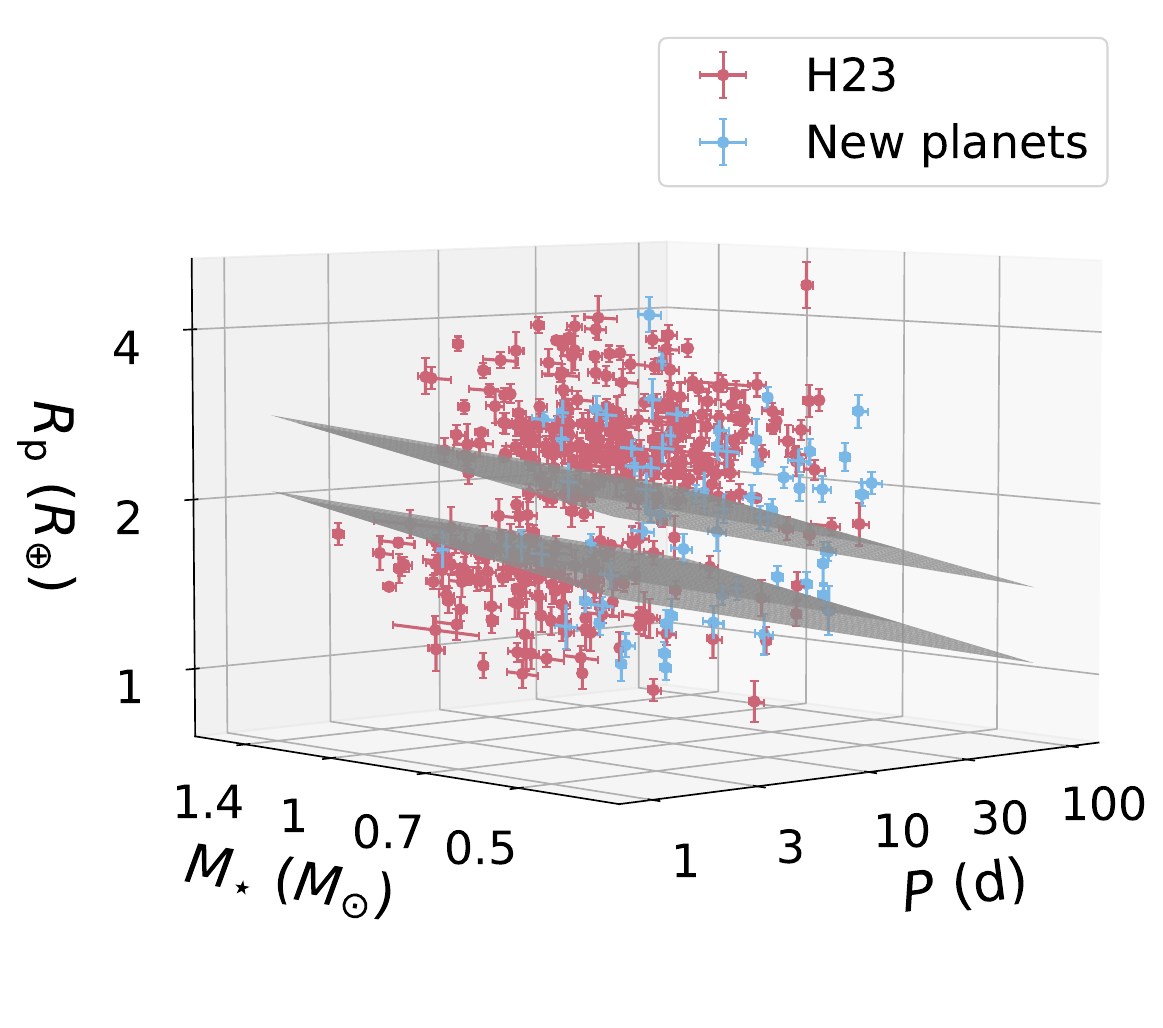}
    \vspace{-8mm}
    \caption{Planetary radius-orbital period-stellar mass plot of the 444 small planets analysed in this work, which includes 375 with parameters from \citet{ho2023deep} (red) and 69 newly refitted (blue) in this work. The radius valley boundary determined in \citet{ho2023deep} is indicated by the grey planes.}
    \label{fig:rpm_allplanets}
\end{figure}

This results in a sample of 69 planets whose properties are homogeneously determined and precisely and accurately known. We then combine these planets with planets with similar characteristics, orbiting higher-mass stars, and include 375 such planets with parameters taken from \citet{ho2023deep}. This total sample of 444 planets is shown in
Fig.~\ref{fig:rpm_allplanets}.

\begin{table*}
    \renewcommand*{\arraystretch}{1.25}
    \centering
    \caption{Table showing MCMC parameter estimates of orbital period ($P$), planetary-to-stellar-radius ratio ($R_{\text{p}}/R_{\star}$), planetary radius ($R_{\text{p}}$), stellar radius ($R_{\star}$), stellar mass ($M_{\star}$), number of transits in the fitted light curve ($N_{\text{tr}}$), of the 72 planets refitted in this work. We convert $R_{\text{p}}/R_{\star}$ into $R_{\text{p}}$ using $R_{\star}$ values from \citet{petigura2022california}. $M_{\star}$ values are also taken from \citet{petigura2022california}. ‘Flag’ refers to whether the planet passes the ﬁlter checks and is included in the smaller subset for further analyses (1 for true and 0 for false). Only the first 5 planets are shown here; the full table is available online in a machine-readable format.}
    \begin{tabular}{ccccccccc}
        \hline 
KOI & Kepler name & $P$ (d) & $R_{\text{p}}/R_{\star}$ & $R_{\text{p}}$ ($R_{\earth}$) & $R_{\star}$ ($R_{\sun}$) & $M_{\star}$ ($M_{\sun}$) & $N_{\text{tr}}$ & Flag \\ 
\hline 
K00156.01 & Kepler-114 c & $8.041329 \pm 0.000005$ & $0.0229 \pm 0.0004$ & $1.80_{-0.06}^{+0.06}$ & $0.72_{-0.02}^{+0.02}$ & $0.73_{-0.03}^{+0.03}$ & 152 & 1 \\ 
K00156.02 & Kepler-114 b & $5.188560 \pm 0.000004$ & $0.0172 \pm 0.0003$ & $1.35_{-0.05}^{+0.05}$ & $0.72_{-0.02}^{+0.02}$ & $0.73_{-0.03}^{+0.03}$ & 234 & 1 \\ 
K00156.03 & Kepler-114 d & $11.776137 \pm 0.000005$ & $0.0364 \pm 0.0005$ & $2.85_{-0.09}^{+0.09}$ & $0.72_{-0.02}^{+0.02}$ & $0.73_{-0.03}^{+0.03}$ & 108 & 1 \\ 
K00222.01 & Kepler-120 b & $6.312501 \pm 0.000005$ & $0.0328 \pm 0.0007$ & $2.54_{-0.09}^{+0.09}$ & $0.71_{-0.02}^{+0.02}$ & $0.70_{-0.03}^{+0.03}$ & 57 & 1 \\ 
K00222.02 & Kepler-120 c & $12.794560 \pm 0.000012$ & $0.0267 \pm 0.0007$ & $2.07_{-0.08}^{+0.08}$ & $0.71_{-0.02}^{+0.02}$ & $0.70_{-0.03}^{+0.03}$ & 27 & 1 \\ 
... & ... & ... & ... & ... & ... & ... & ... & ... \\ 
\hline
    \end{tabular}
    \label{tab:planet_params}
\end{table*}

\begin{figure}
    \centering    \includegraphics[width=\columnwidth]{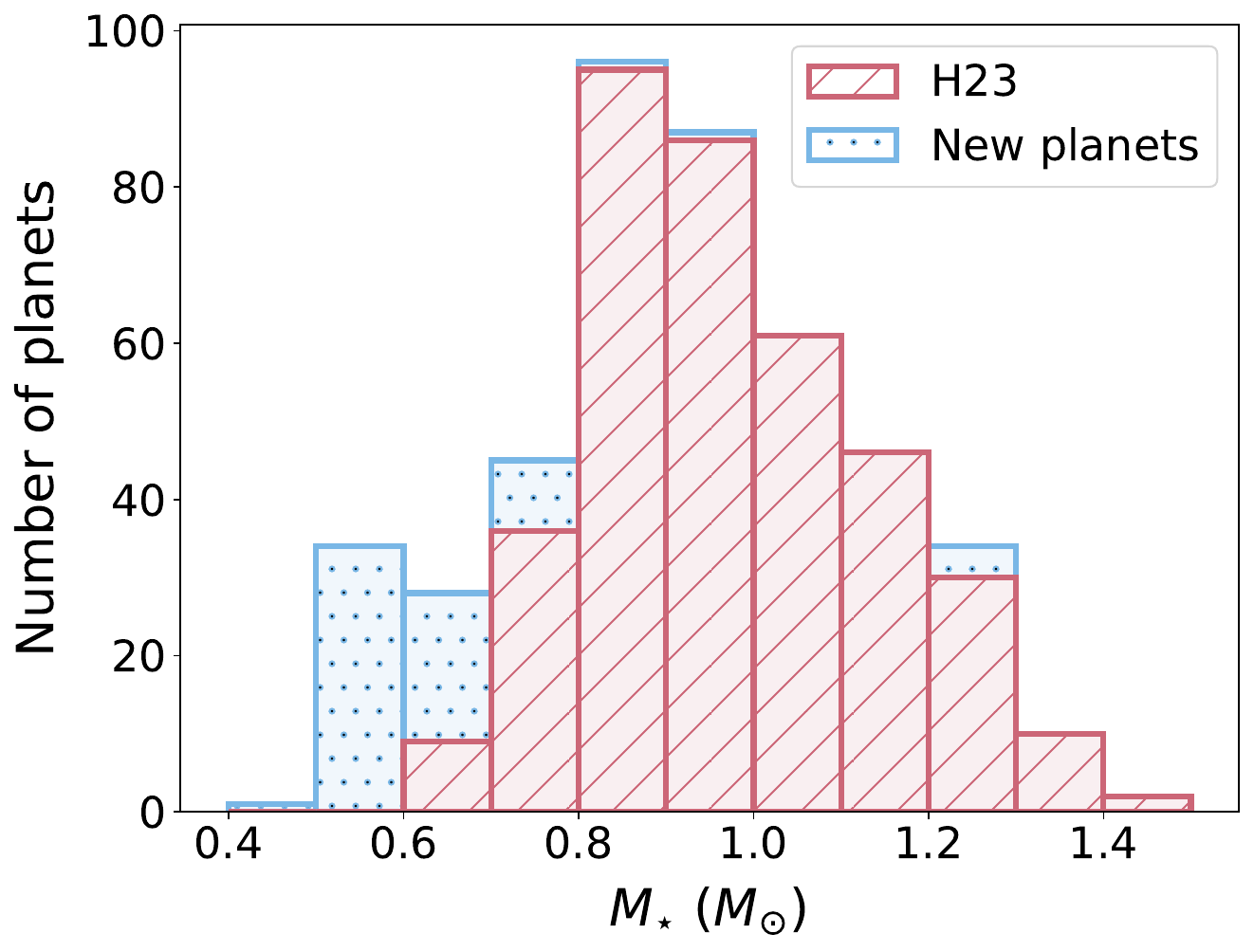}
    \caption{Stacked histogram comparing the distribution of the stellar mass of the sample of the new planets used in this work, and planets in \citet{ho2023deep}.}
    \label{fig:hist_ms_compare}
\end{figure}

\begin{figure*}
    \centering
    \begin{tabular}{cc}
        \includegraphics[width=0.48\linewidth]{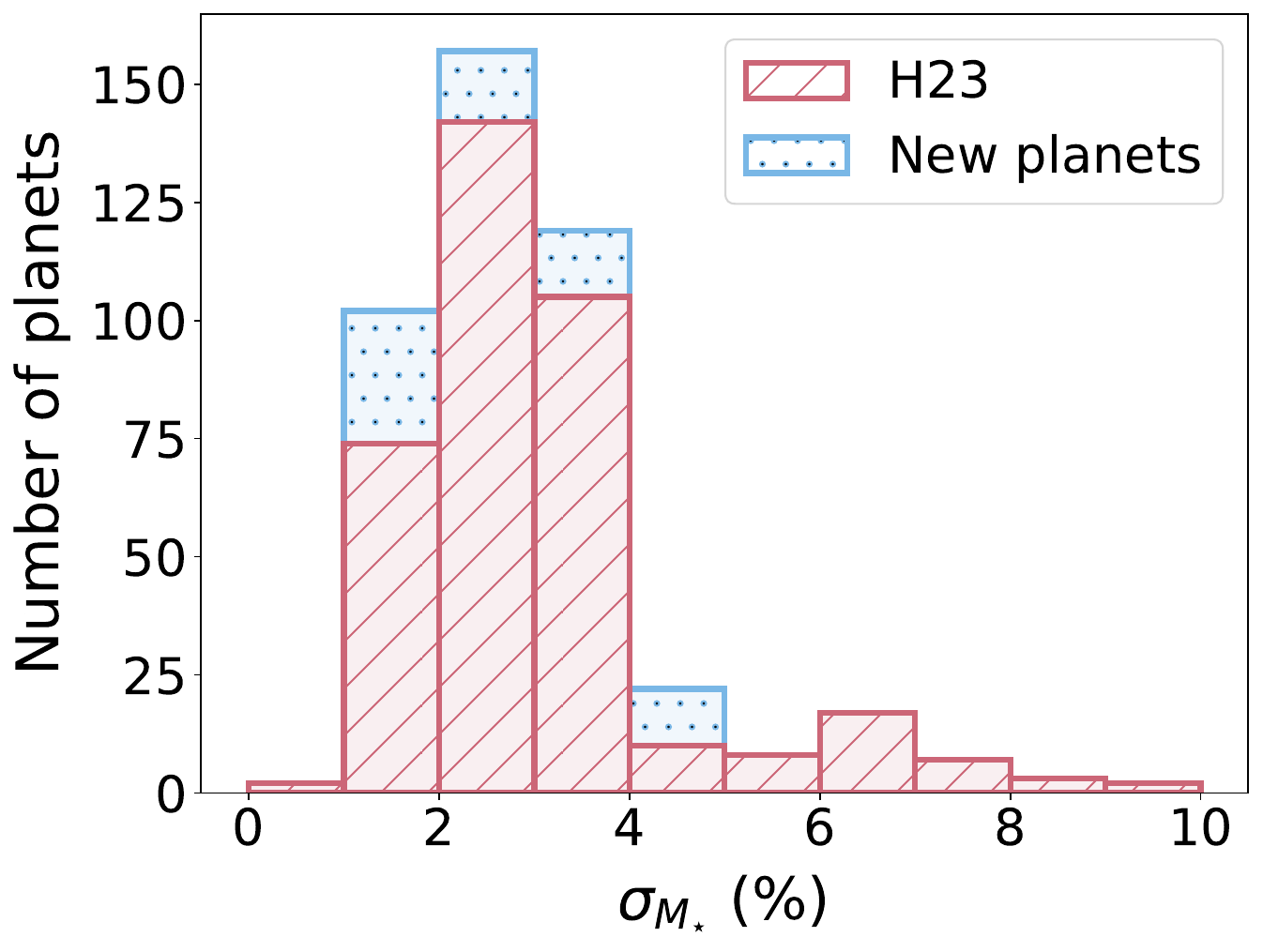} & \includegraphics[width=0.48\linewidth]{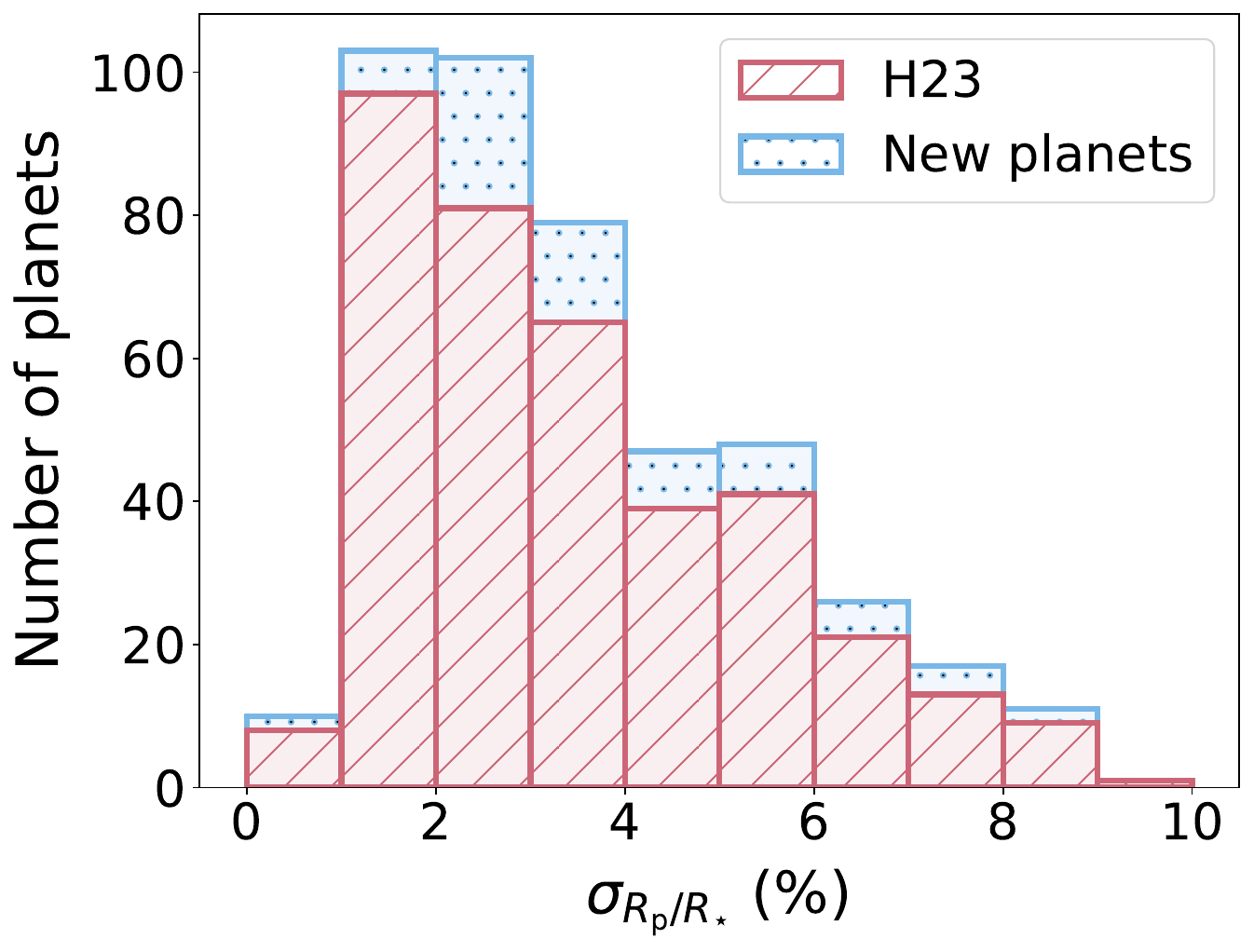} \\\includegraphics[width=0.48\linewidth]{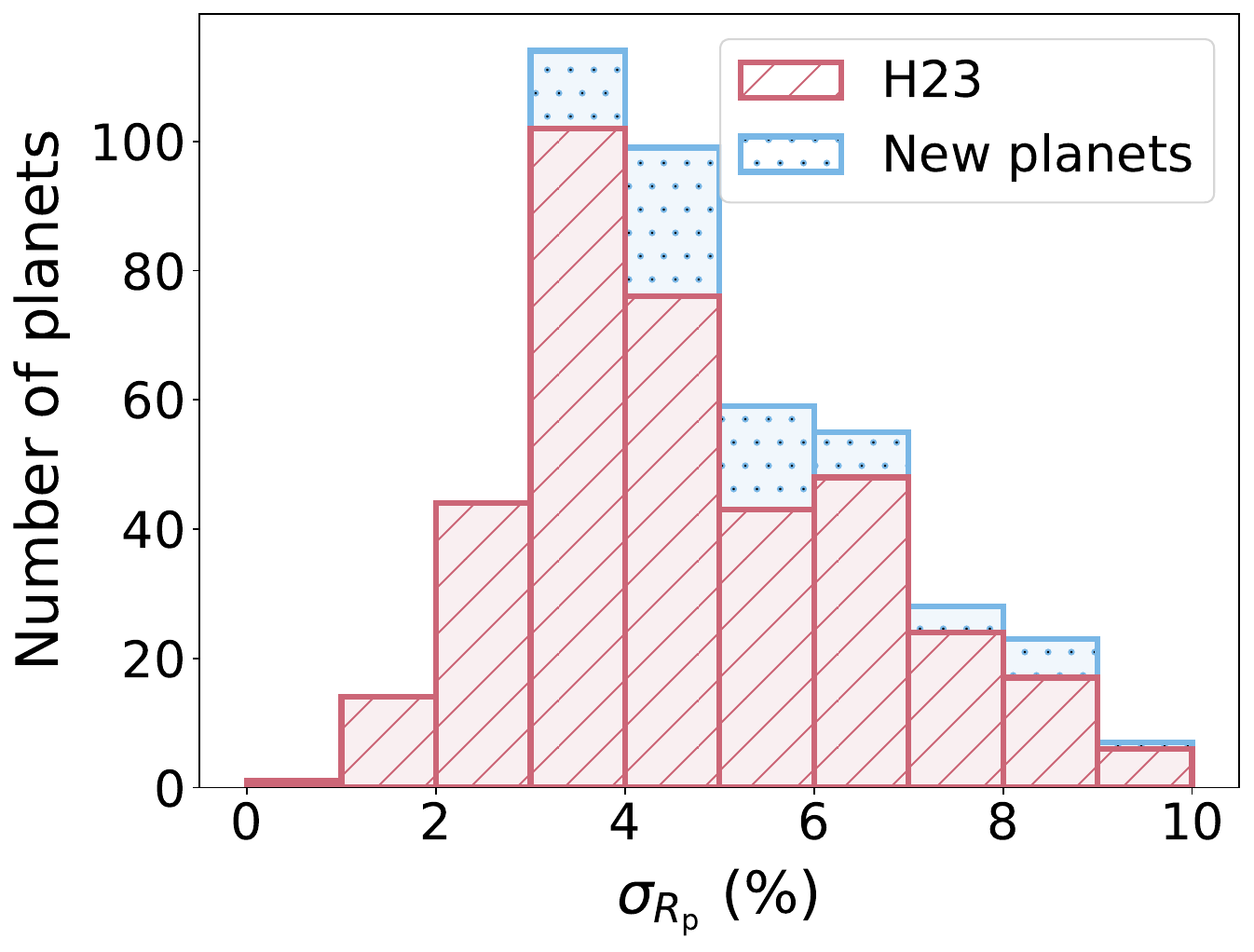} &
        \includegraphics[width=0.48\linewidth]{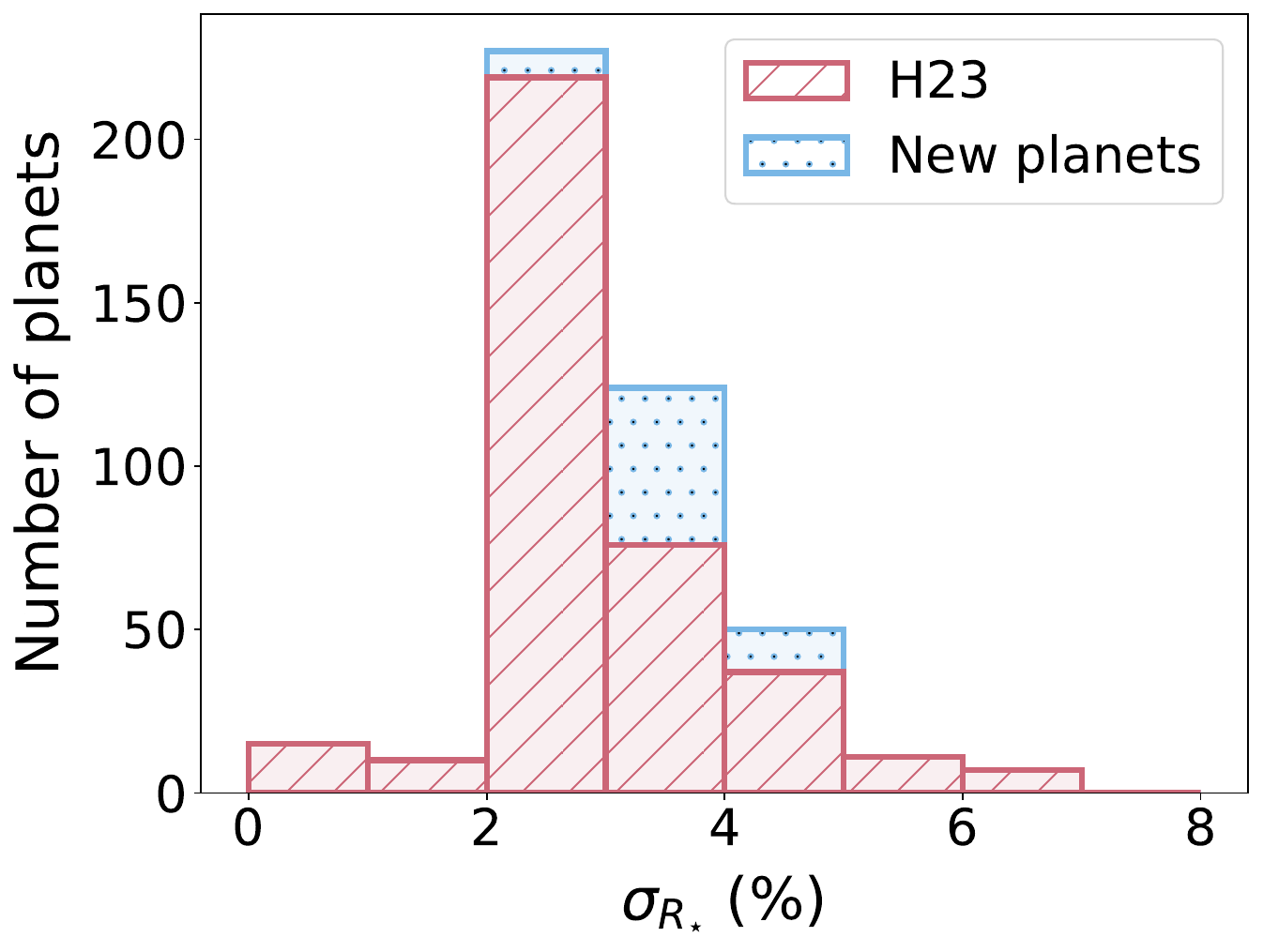}
    \end{tabular}
    \caption{Same as Fig.~\ref{fig:hist_ms_compare}, but for percentage uncertainty in $M_{\star}$ ($\sigma_{M_{\star}}$), planet-to-star radius ratio ($\sigma_{R_{\text{p}}/R_{\star}}$), planetary radius ($\sigma_{R_{\text{p}}}$), and stellar radius ($\sigma_{R_{\star}}$).}
    \label{fig:hist_others_compare}
\end{figure*}

The sample of planets with properties newly determined in this work provides an important complement to the sample studied by \cite{ho2023deep}. In Fig.~\ref{fig:hist_ms_compare}, we show both samples as a function of stellar mass, highlighting how these 69 new planets significantly extend the homogeneous sample around low-mass M dwarfs. In Fig.~\ref{fig:hist_others_compare}, we compare the noise properties of both samples, showing that the uncertainties in planet-to-star radius ratio, planet radius, stellar mass and radius, are all broadly similar.

\section{Radius valley of low-mass stars} \label{sect:results}

\subsection{Shallower observed radius valley for lower mass stars} \label{subsect:results_shallow}
\begin{figure*}
    \centering
    \begin{tabular}{cc}
         \includegraphics[width=\columnwidth]{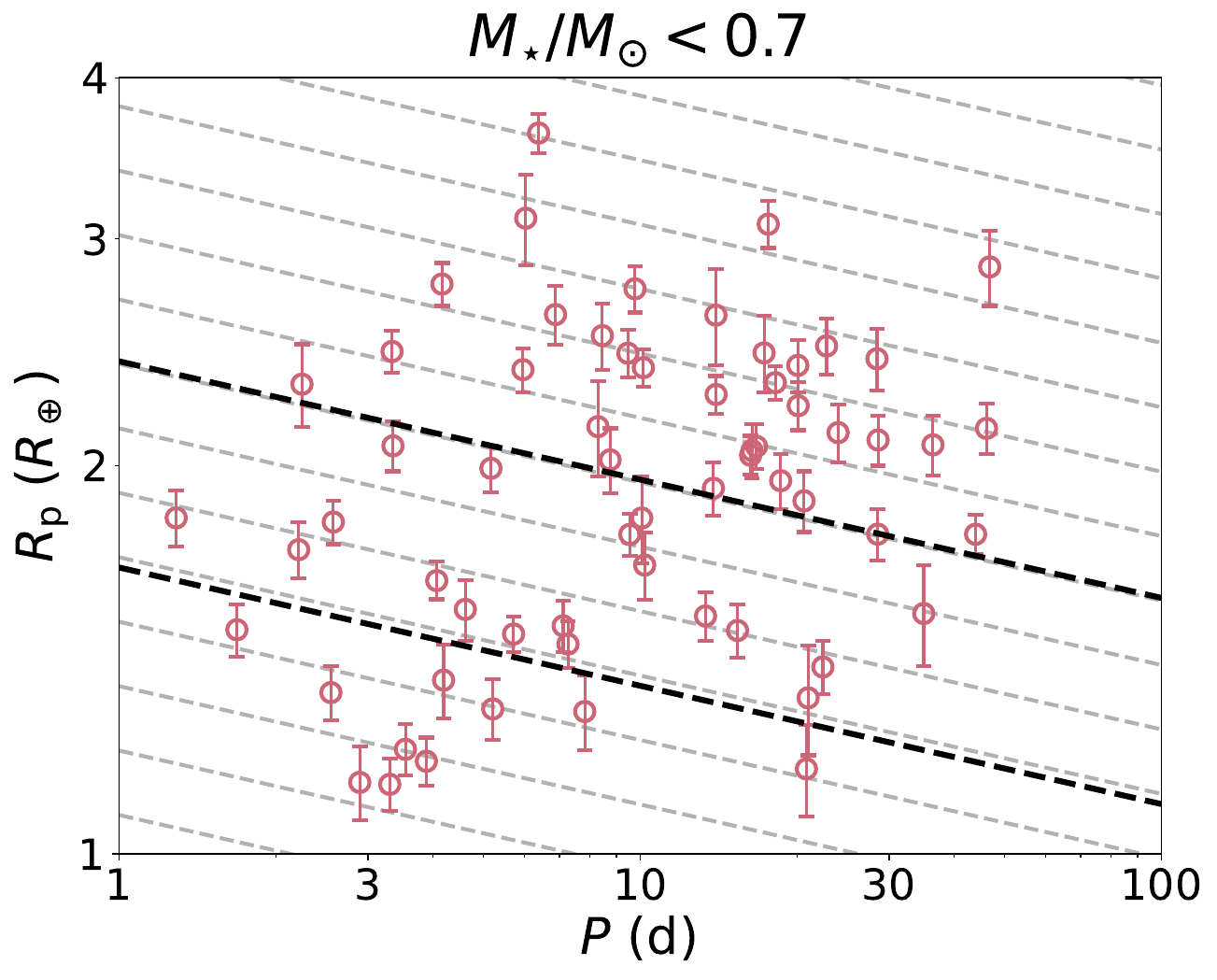} & \includegraphics[width=\columnwidth]{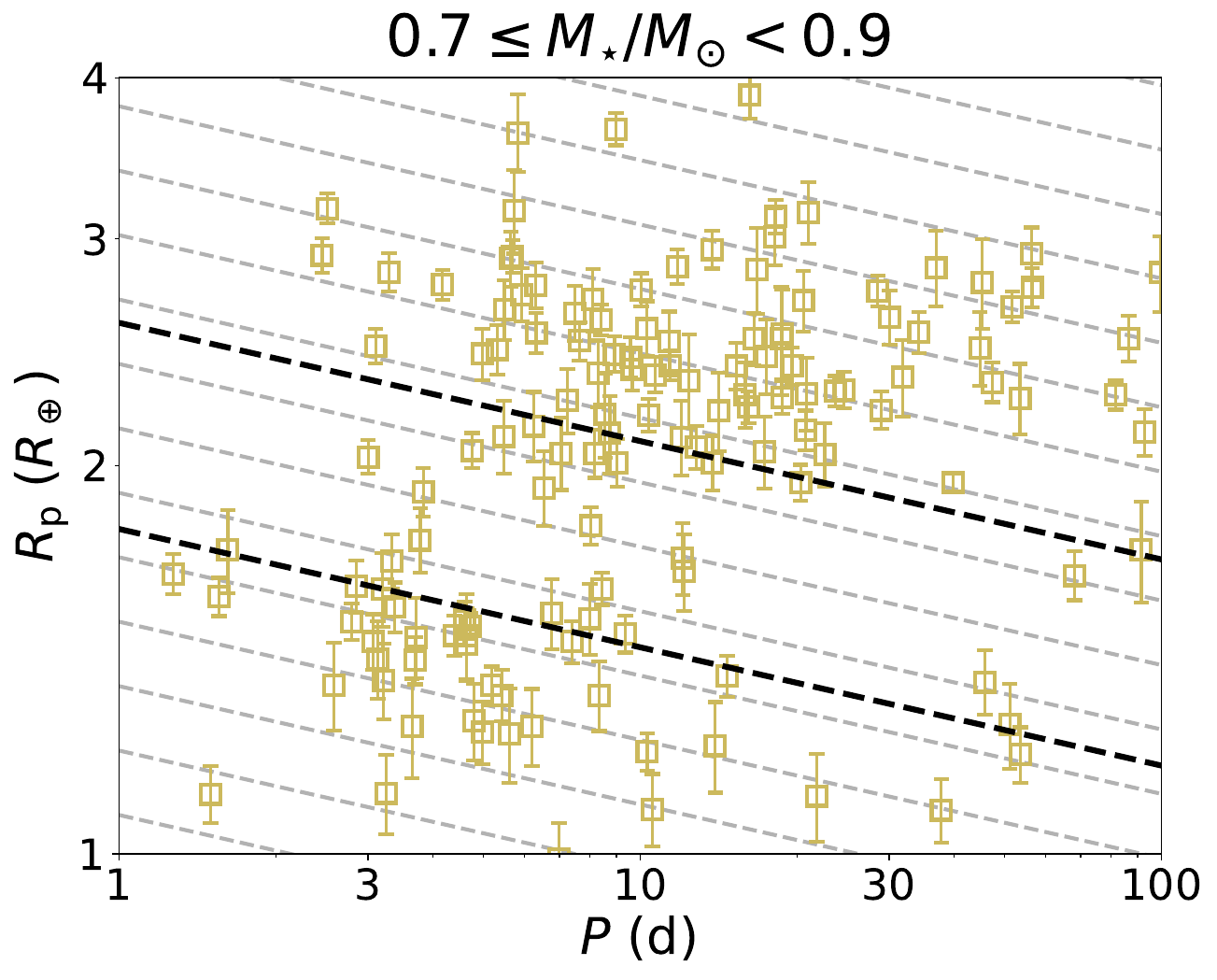} \\
         \includegraphics[width=\columnwidth]{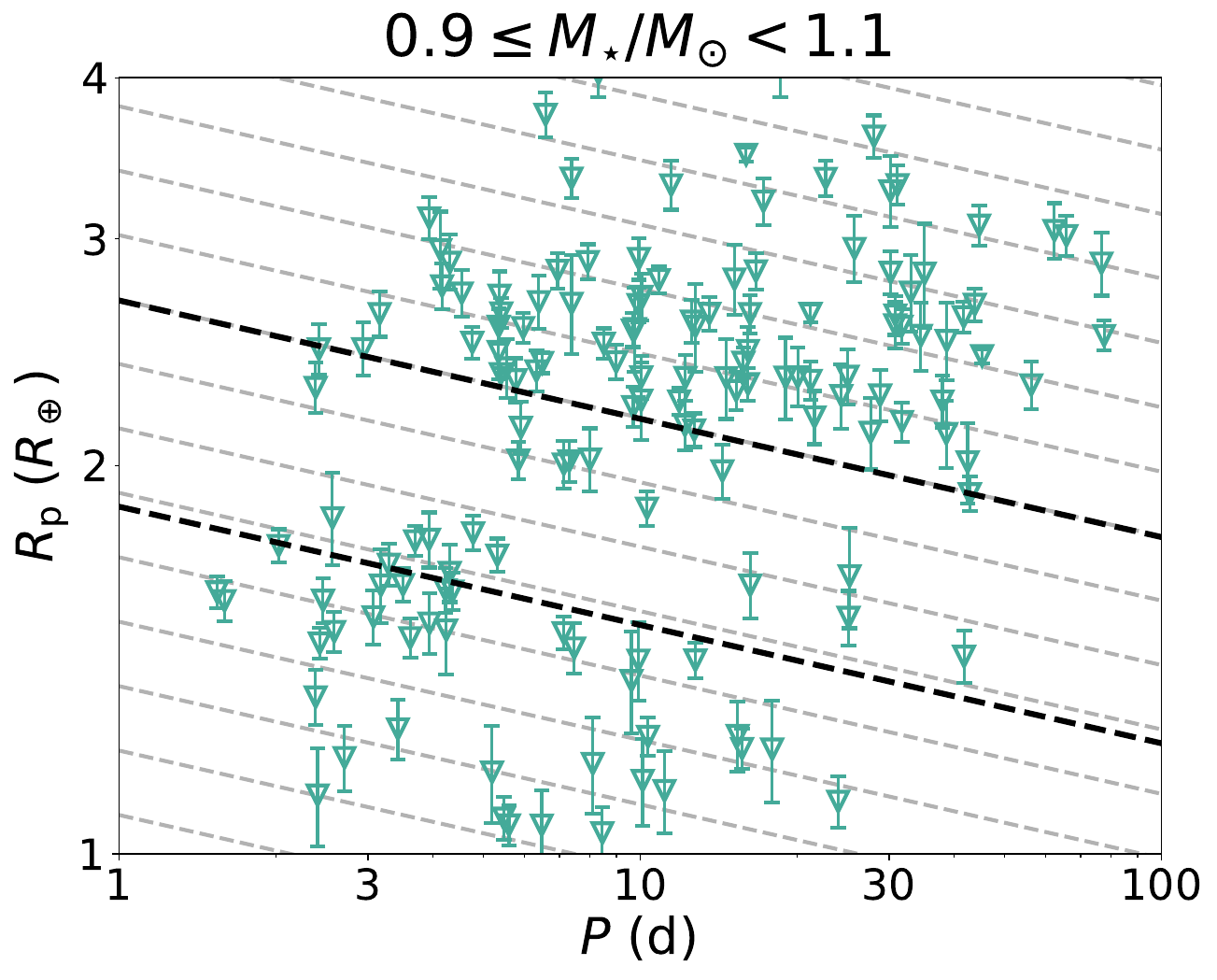} & \includegraphics[width=\columnwidth]{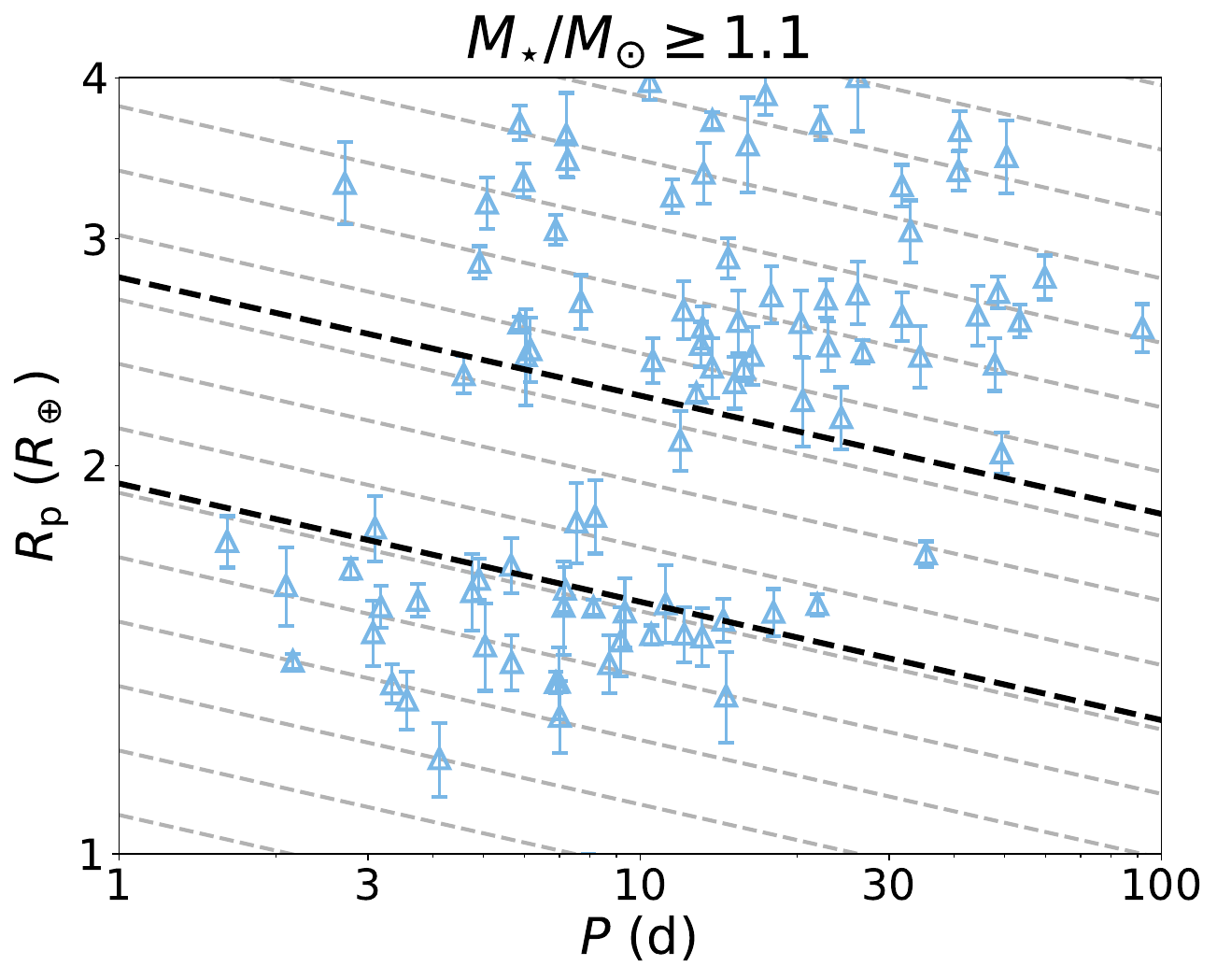}
    \end{tabular}
    \caption{Radius-orbital period plot of planets analysed in this work, separated by host star mass. The black dashed lines indicate the radius valley boundary as defined by the parameters in \citet{ho2023deep}. The grey dashed lines divide the plot into multiple orbital-period dependent bins, which are then used to plot the adjusted histogram in Fig.~\ref{fig:tilthist_bymstar}.}
    \label{fig:rp_obs_bymstar}
\end{figure*}

We now investigate the properties of the radius valley as a function of stellar mass. As a starting point, we divide the sample into four mass bins: $M_{\star}/M_{\sun} < 0.7$, $0.7 \leq M_{\star}/M_{\sun} < 0.9$, $0.9 \leq M_{\star}/M_{\sun} < 1.1$, and $M_{\star}/M_{\sun} \geq 1.1$.  The resulting sample in those four stellar mass bins is shown as a function of planet radius and orbital period in Fig.~\ref{fig:rp_obs_bymstar}. On the same figure, we also show the boundaries of the radius valley, according to the 3-dimensional equation determined in \citet{ho2023deep}, which is defined by a support-vector machine (SVM) separating the planet population into two groups:
\begin{equation}
    \log_{10}\left({R_{\text{p}}/R_{\earth}}\right) = A\log_{10}\left({P/\text{d}}\right) + B\log_{10}\left({M_{\star}/M_{\sun}}\right) + C
    \label{eq:rv_3d_rpm}
\end{equation}
with $A=-0.09^{+0.02}_{-0.03}$, $B=0.21^{+0.06}_{-0.07}$, $C=0.35^{+0.02}_{-0.03}$, and setting $M_{\star}$ to the median of the host stars masses within each mass range. We take ${C_\text{upper}} = 0.42$ and ${C_\text{lower}} = 0.29$ from \citet{ho2023deep}, which are determined by the parallel lines passing through the supporting vectors. Here, we have plotted the valley as determined by \citet{ho2023deep} --- without the contribution from the lower-mass stars. When including the latter and refitting equation~\ref{eq:rv_3d_rpm}, we find $A = -0.10^{+0.02}_{-0.03}$, $B = 0.17 ^{+0.14}_{-0.09}$, $C = 0.36^{+0.02}_{-0.02}$, which are in agreement with \citet{ho2023deep} well within $1\sigma$. We also test whether the radius valley slope differs for individual stellar mass bins. We fit a 2-dimensional SVM in radius-period space for each stellar mass range. We find a slope of $-0.15^{+0.08}_{-0.18}$, $-0.10^{+0.02}_{-0.04}$, $-0.10^{+0.04}_{-0.10}$, and $-0.11^{+0.05}_{-0.05}$ for $M_{\star}/M_{\sun} < 0.7$, $0.7 \leq M_{\star}/M_{\sun} < 0.9$, $0.9 \leq M_{\star}/M_{\sun} < 1.1$, and $M_{\star}/M_{\sun} \geq 1.1$ respectively. These slopes are also in agreement with each other well within $1\sigma$. As one of the main goals of this work is precisely to test whether the radius valley has the same physical origin around low-mass stars as around their higher-mass counterparts, we choose here not to use these updated radius valley parameters.

\begin{figure*}
    \centering
    \includegraphics[width=\linewidth]{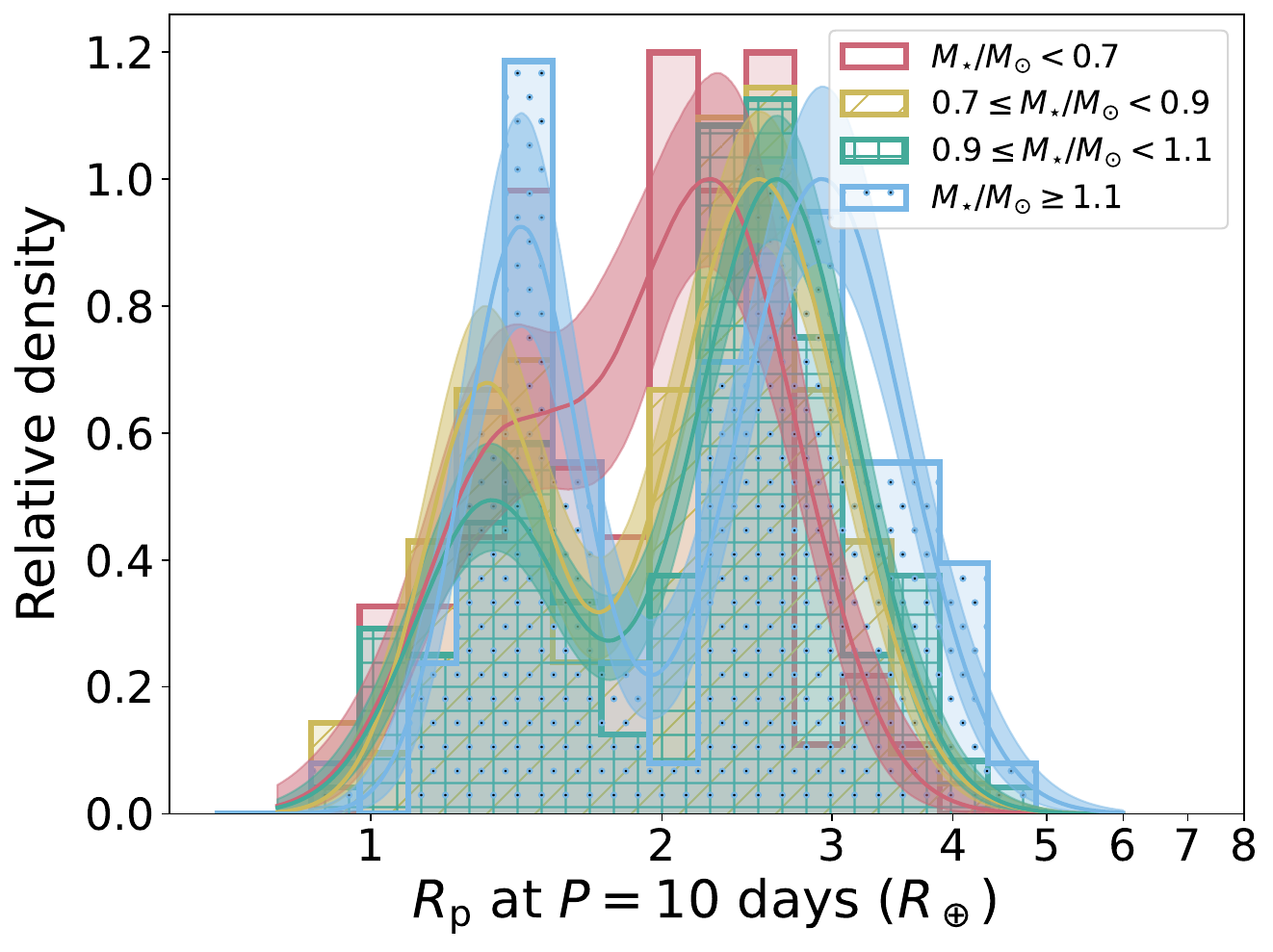}
    \caption{Histogram of $R_{\text{p}}$, adjusted to an equivalent radii at P = 10 d according to the valley slope determined in \citet{ho2023deep} ($m = -0.09$), for planets of different host star masses. The histograms are normalised such that the peaks of the sub-Neptune curves have a relative density of 1. The lines and shaded regions show the median and $1\sigma$ confidence interval of the fitted curve respectively.}
    \label{fig:tilthist_bymstar}
\end{figure*}

Visually inspecting the distribution of planet radii and orbital periods in Fig.~\ref{fig:rp_obs_bymstar}, we see that for higher mass stars, there are fewer planets lying within the boundaries of the radius valley. 
Since the radius valley is orbital-period dependent \citep[e.g.][]{vaneylen2018asteroseismic, ho2023deep}, we divide the plot into multiple tilted bins, as shown in Fig.~\ref{fig:rp_obs_bymstar}, and following the procedure in \citet{ho2023deep}, shift the planets along the slope of the valley to an equivalent radii at $P = 10$ d ($R_\text{p,10}$). We then plot a histogram of these equivalent radii as shown in Fig.~\ref{fig:tilthist_bymstar}, separated by different stellar mass ranges. Again, following the procedure in \citet{ho2023deep}, we fit the histogram with a curve determined using a Gaussian mixture model with two components, which is independent of the width and boundaries of the histogram bins, and fixes the bimodal distribution of small planets. This equation is given by
\begin{equation}
\begin{split}
    \lambda(R_\text{p,10}) = w_1 \times \frac{1}{\sqrt{2\pi}\sigma_1} \exp \left({\frac{-\left (R_\text{p,10} - \mu_1 \right)^2}{2\sigma_1^2}} \right) \\ + w_2 \times \frac{1}{\sqrt{2\pi}\sigma_2} \exp \left({\frac{-\left (R_\text{p,10} - \mu_2 \right)^2}{2\sigma_2^2}} \right),
\end{split}
    \label{eq:gmm}
\end{equation}
with $w_1 + w_2 = 1$. To account for measurement uncertainties, we follow the method used in \citet{Rogers2021Unveiling} and fit the Gaussian mixture model as an inhomogeneous Poisson point process. The likelihood function of observing an ensemble of planets at $\theta = \{ R_\text{p,10,1}, ..., R_{\text{p},\text{10},N_{\text{pl}}} \}$ is given by
\begin{equation}
    L(\theta) = \mathrm{e}^{-\Lambda} \frac{\Lambda^{N_\text{pl}}}{N_{\text{pl}}!} \prod_{i=1}^{N_{\text{pl}}} \frac{\lambda \left(R_{\text{p}, \text{10}, i}\right)}{\Lambda}
\end{equation}
where $\Lambda = A \int \lambda \left(R_\text{p,10}\right) \, \mathrm{d} R_\text{p,10}$ is the underlying occurrence rate, with $A$ being an arbitrary constant which we also fit for, $N_{\text{pl}}$ is the number of observed planets, and $\lambda \left(R_{\text{p},\text{10},i}\right)/\Lambda$ is the probability density of observing planet $i$ at $R_{\text{p},\text{10},i}$. We infer the posteriors for $A$, $\mu_1$, $\mu_2$, $\sigma_1$, $\sigma_2$, and $w_1$ by inputting the log-likelihood, $\ln{(L)}$, into a Markov Chain Monte Carlo (MCMC) algorithm, using the Python package \texttt{emcee} \citep{foremanmackey2013emcee} with 100 walkers and 5000 steps. The median and $\pm 1\sigma$ uncertainties are shown on the fitted curves. The posterior estimates of $A$, $\mu_1$, $\mu_2$, $\sigma_1$, $\sigma_2$, $w_1$, and $w_2$ are reported in Table~\ref{tab:poisson_posteriors}.

\begin{table*}
\renewcommand*{\arraystretch}{1.25}
    \centering
    \caption{Posterior estimates of parameters from fitting the Gaussian mixture model as an inhomogeneous Poisson point process.}
    \begin{tabular}{cccccccc}
    \hline
        $M_{\star}$ ($M_{\sun}$) & $A$ & $\mu_1$ & $\mu_2$ & $\sigma_1$ & $\sigma_2$ & $w_1$ & $w_2$ \\
    \hline
        $M_{\star}<0.7$ & $0.70^{+1.17}_{-0.53}$ & $0.13^{+0.03}_{-0.03}$ & $0.35^{+0.02}_{-0.03}$ & $0.08^{+0.03}_{-0.02}$ & $0.09^{+0.02}_{-0.02}$ & $0.35^{+0.12}_{-0.14}$ & $0.65^{+0.14}_{-0.12}$ \\
        $0.7 \leq M_{\star} < 0.9$ & $0.68^{+1.15}_{-0.51}$ & $0.12^{+0.01}_{-0.01}$ & $0.40^{+0.01}_{-0.01}$ & $0.06^{+0.01}_{-0.01}$ & $0.09^{+0.01}_{-0.01}$ & $0.33^{+0.05}_{-0.04}$ & $0.67^{+0.04}_{-0.05}$ \\
        $0.9 \leq M_{\star} < 1.1$ & $0.70^{+1.15}_{-0.52}$ & $0.12^{+0.02}_{-0.01}$ & $0.42^{+0.01}_{-0.01}$ & $0.08^{+0.01}_{-0.01}$ & $0.08^{+0.01}_{-0.01}$ & $0.32^{+0.05}_{-0.04}$ & $0.68^{+0.04}_{-0.05}$ \\
        $1.1 \leq M_{\star} < 1.3$ & $0.70^{+1.14}_{-0.52}$ & $0.16^{+0.01}_{-0.01}$ & $0.47^{+0.01}_{-0.01}$ & $0.06^{+0.01}_{-0.01}$ & $0.09^{+0.01}_{-0.01}$ & $0.39^{+0.05}_{-0.05}$ & $0.61^{+0.05}_{-0.05}$ \\
    \hline
    \end{tabular}
    \label{tab:poisson_posteriors}
\end{table*}

\begin{figure}
    \centering
    \includegraphics[width=\columnwidth]{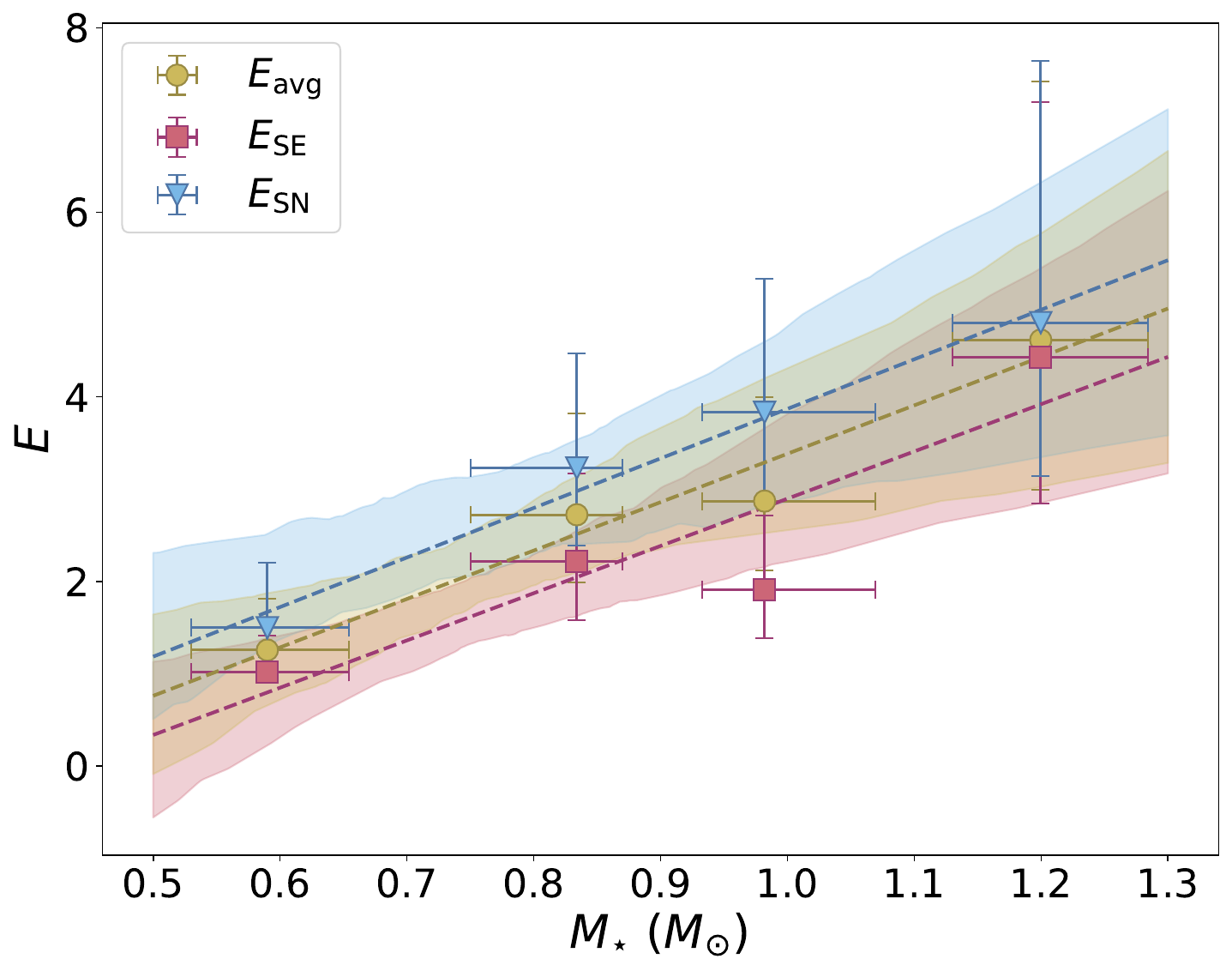}
    \caption{Plot of the $E$ metric to measure the depth of the radius valley from equations~\ref{eq:e_avg}-\ref{eq:e_sn}, as a function of host star mass. The shaded regions represent the $\pm 1 \sigma$ confidence intervals of the linear fits.}
    \label{fig:E_bymstar}
\end{figure}

We observe two features in this histogram. The first one is that the radius valley gets deeper for higher stellar mass. We use the metric defined in \citet{ho2023deep}:
\begin{equation}
    E_{\text{avg}} = (E_{\text{SE}} + E_{\text{SN}})/2
    \label{eq:e_avg}
\end{equation}
to compare the depth of the radius valley, with
\begin{equation}
    E_{\text{SE}} = N_{\text{super-Earth, peak}}/N_{\text{valley}}
    \label{eq:e_se}
\end{equation}
and
\begin{equation}
    E_{\text{SN}} = N_{\text{sub-Neptune, peak}}/N_{\text{valley}}
    \label{eq:e_sn}
\end{equation}
where $N_{\text{super-Earth, peak}}$ and $N_{\text{sub-Neptune, peak}}$ refer to the number of planets at the super-Earth and sub-Neptune Gaussian peaks, taken from the median values of the fitted Gaussian mixture model (equation~\ref{eq:gmm}), respectively, and $N_{\text{valley}}$ is the lowest number of planets between the two Gaussian peaks. The resulting $E$ values for different stellar mass ranges are reported in Table~\ref{tab:E_mstar} and plotted in Fig.~\ref{fig:E_bymstar}. From Fig.~\ref{fig:E_bymstar}, we see that both $E_{\text{avg}}$ and $E_{\text{SN}}$ strictly increases with stellar mass,  whereas $E_{\text{SE}}$ generally increases with stellar mass except at the transition at around $0.9M_{\sun}$; this may however be due to the decreasing detection efficiency of small planets for higher mass stars. We perform a linear fit to the $E$ values as a function of stellar mass. The resulting slopes show a clear increasing trend with a slope of $5.36^{+3.19}_{-3.23}$ for $E_{\text{avg}}$, $4.79^{+3.44}_{-2.19}$ for $E_{\text{SE}}$, and $4.91^{+3.05}_{-3.37}$ for $E_{\text{SN}}$. The p-values from the Wald Test with t-distribution of the test statistic, as implemented in \texttt{scipy.stats.linregress} \citep{2020SciPy-NMeth}, for the linear fit are 0.02, 0.09 and 0.01, respectively, indicating a strong rejection of the null hypothesis of no linear correlation between the two parameters. We do not expect that observational biases could reproduce these trends. The planet detection efficiency should decrease with increasing stellar radius (and therefore mass), due to the transit depth scaling inversely with the square of stellar radius, i.e. $\delta \propto \left(R_{\text{p}}/R_{\star}\right)^2$. Although \textit{Kepler} is less sensitive to cooler M-dwarfs, their detection efficiency is comparable to their FGK counterparts \citep[$\sim 92\%$ compared to $\sim 96\%$, ][]{christiansen2020measuring}. Hence, the deeper radius valley for higher-mass stars should not be caused by undetected planets. The relative uncertainty on stellar radius is also similar for low-mass stars as for their higher-mass counterparts (see Figure~\ref{fig:hist_others_compare}).

\begin{table}
\renewcommand*{\arraystretch}{1.25}
    \centering
    \caption{$E$ values of the radius valley for planets with different host star masses. The $E$ metrics are calculated from equations~\ref{eq:e_avg}-\ref{eq:e_sn}.}
    \begin{tabular}{ccccc}
    \hline
        $M_{\star}$ $(M_{\sun})$ & $N_{\text{planets}}$ & $E_{\text{avg}}$ & $E_{\text{SE}}$ & $E_{\text{SN}}$ \\
    \hline
        $M_{\star}<0.7$ & 63 & $1.25^{+0.55}_{-0.17}$ & $1.01^{+0.40}_{-0.01}$ & $1.50^{+0.71}_{-0.32}$ \\
        $0.7 \leq M_{\star} < 0.9$ & 141 & $2.72^{+1.10}_{-0.74}$ & $2.22^{+0.95}_{-0.63}$ & $3.23^{+1.25}_{-0.84}$ \\
        $0.9 \leq M_{\star} < 1.1$ & 148 & $2.87^{+1.13}_{-0.76}$ & $1.91^{+0.81}_{-0.53}$ & $3.83^{+1.45}_{-0.98}$ \\
        $M_{\star} \geq 1.1$ & 92 & $4.62^{+2.80}_{-1.62}$ & $4.43^{+2.77}_{-1.58}$ & $4.80^{+2.84}_{-1.66}$ \\
    \hline
    \end{tabular}
    \label{tab:E_mstar}
\end{table}

\begin{figure}
    \centering
    \includegraphics[width=\columnwidth]{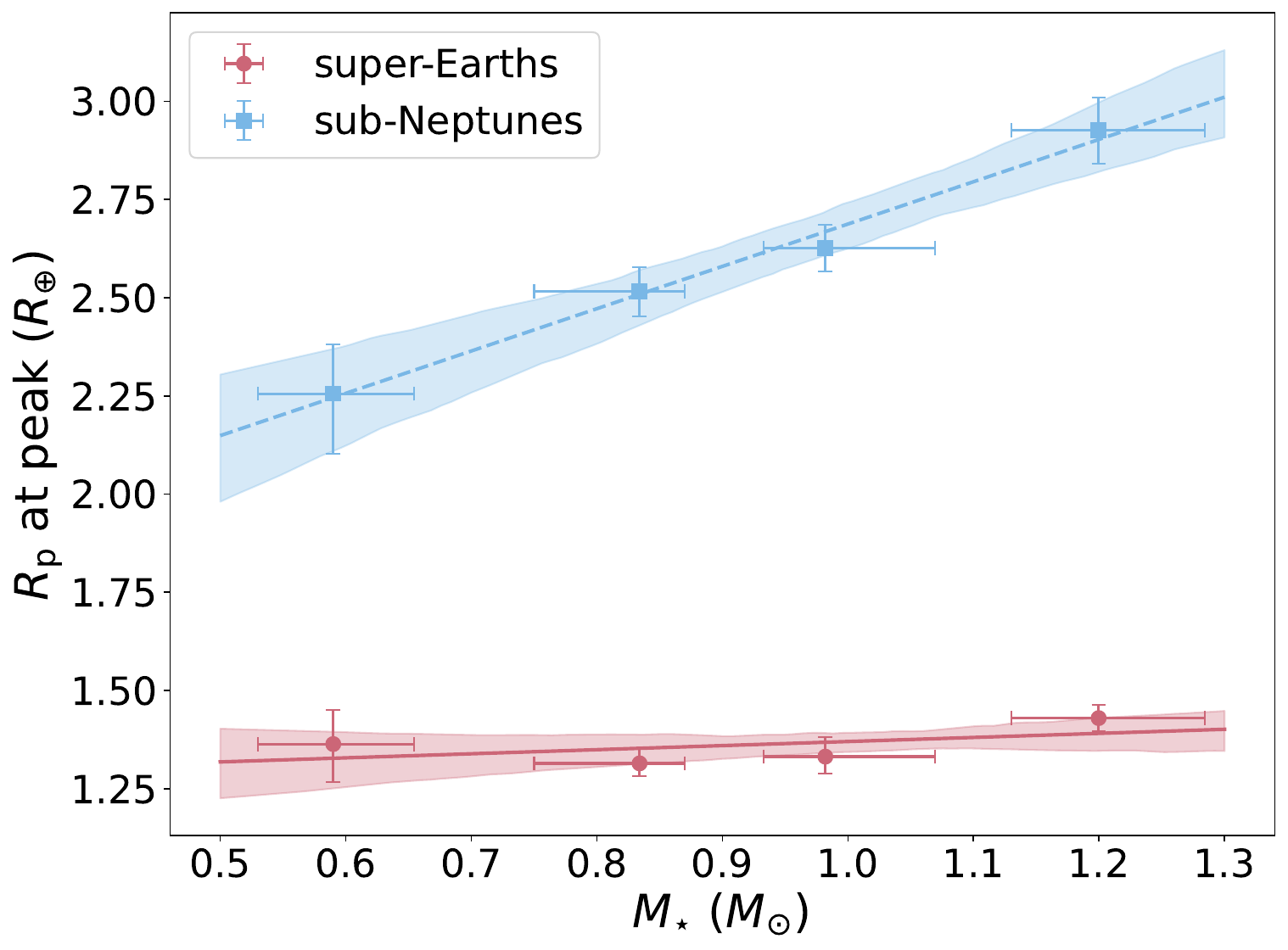}
    \caption{Plot of the planetary radius at the super-Earth and sub-Neptune histogram peaks from Fig.~\ref{fig:tilthist_bymstar} as a function of host star mass. The shaded regions represent the $\pm 1 \sigma$ confidence intervals of the linear fits.}
    \label{fig:Rp_peak_bymstar}
\end{figure}

The second feature we observe is that the sub-Neptune peaks shift to higher $R_{\text{p}}$ for higher mass stars, whereas the super-Earth peaks do not increase as much. Fig.~\ref{fig:Rp_peak_bymstar} shows the radius of the super-Earth and sub-Neptune peaks as a function of stellar mass. Again, we perform a linear fit to the planet radii at the two peaks as a function of stellar mass, and find $R_\text{p,peak} = 0.13^{+0.16}_{-0.19} M_{\star} + 1.25^{+0.19}_{-0.17}$ for super-Earths, and $R_\text{p,peak} = 1.09^{+1.30}_{-0.29} M_{\star} + 1.58^{+0.65}_{-0.31}$ for sub-Neptunes. The p-values of 0.45 and 0.005, respectively indicate a strong rejection of the no linear correlation hypothesis for sub-Neptunes, but not for the case of super-Earths. This effect is also confirmed by other observational studies \citep[e.g.][]{petigura2022california}, which analysed a similar sample of planets, albeit using \textit{Kepler} long-cadence observations and different transit modelling than the ones used here.

\subsection{Photoevaporation models predict a narrower radius valley for lower mass stars} \label{subsect:results_model}

\citet{owen2024mapping} suggested that although the observed super-Earth population likely contains significant fractions of planets that were stripped by both photoevaporation and core-powered mass-loss, photoevaporation was likely responsible for the final atmospheric stripping of planets located at the edge of the radius valley today, determining the final features observed today. Therefore, we only perform our subsequent comparison between observations and photoevaporation models.

We model planets undergoing thermal cooling and photoevaporative mass loss of their hydrogen-dominated atmospheres with the semi-analytic framework of \citet{Rogers2021Unveiling} and \citet{rogers2023conclusive}. In doing so, we also adopt the underlying planet distributions inferred in \citet{Rogers2021Unveiling} for planetary core masses, initial atmospheric mass fractions, orbital periods, as well as an Earth-like core density for all planets of $5.5 \text{ g cm}^{-3}$. Changing these distributions, which come from statistical inference between the photoevaporation model and \textit{Kepler} data, will inevitably change the population of planets that might be observed under such conditions. This highlights an important approach in this work, in that, we wish to compare \textit{Kepler} data with a realisation of the photoevaporation model that fits the observations as well as reasonably possible. Inevitably, there are multiple parameters within our evolution model that can be tuned within reasonable limits to fit the observations more accurately. However, if inconsistencies still exist between the model and observations, even after tuning, the robustness of the result increases. One such example is that, under the photoevaporation model, the predicted slope of the radius valley as a function of the orbital period is altered with different assumptions in the escape efficiency $\eta$ \citep{owen2017evaporation}, which allows one to calculate mass loss rates:
\begin{equation} \label{eq:energy-limited-Mdot}
     \dot{M} = \eta \, \frac{\pi R_{\text{p}}^3 F_\text{XUV}}{G M_{\text{p}}},
\end{equation} 
where $F_\text{XUV}$ is the incident stellar XUV flux \citep[following][]{rogers2021photoevaporation}, $G$ is the gravitational constant, and  $R_{\text{p}}$ and $M_{\text{p}}$ are the planet's radius and mass. Typically, the efficiencies are parameterised as a function of the planet's escape velocity:
\begin{equation}
     \eta = \eta_0 \; \bigg ( \frac{v_\text{esc}}{23 \text{ km s}^{-1}} \bigg )^{\alpha_\eta}.
\end{equation}
To begin, we find a pair of $\eta_0$ and $\alpha_\eta$ that approximately match the slope of the observed valley. In practice, we achieve this by finding the pair of parameters that minimise the number of \textit{observed} planets inside the \textit{model's} valley. We do this only in the highest stellar mass bin ($1.1 \leq M_{\star}/M_{\sun} < 1.3$) since the photoevaporation models have only ever been fit to FGK stellar hosts, for which an accurate fit has been shown to exist \citep{Rogers2021Unveiling}. Therefore, we fix the models to attain this consistent fit at high stellar mass, which then yields an extrapolation of the model to lower stellar masses. The specific values were found to be $\eta_0 = 0.24$ and $\alpha_\eta=-1.11$, consistent with the inference of \citet{Rogers2021Unveiling} and hydrodynamic simulations of atmospheric escape.

\begin{figure*}
    \centering
    \begin{tabular}{cc}
        \includegraphics[width=\columnwidth]{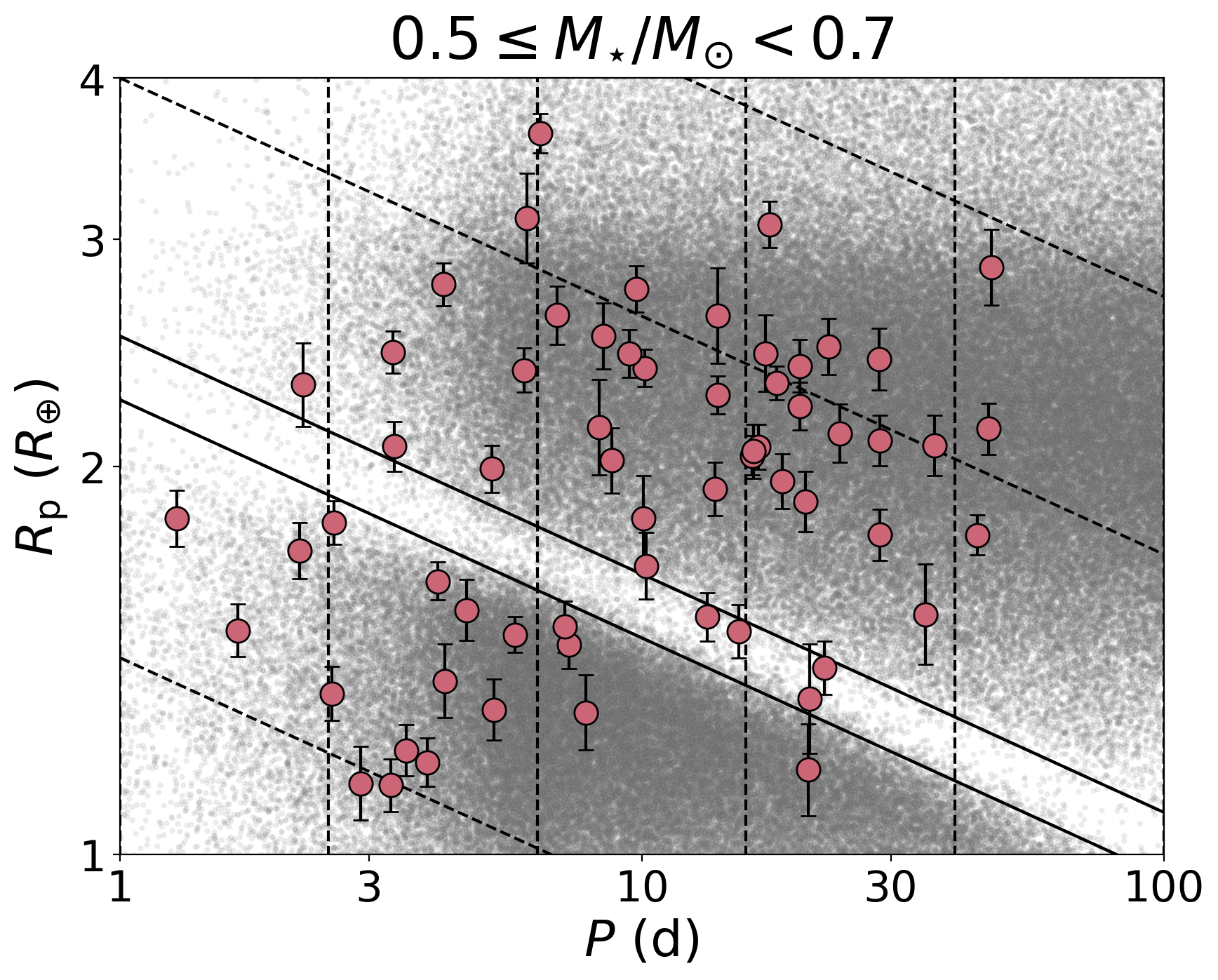} & \includegraphics[width=\columnwidth]{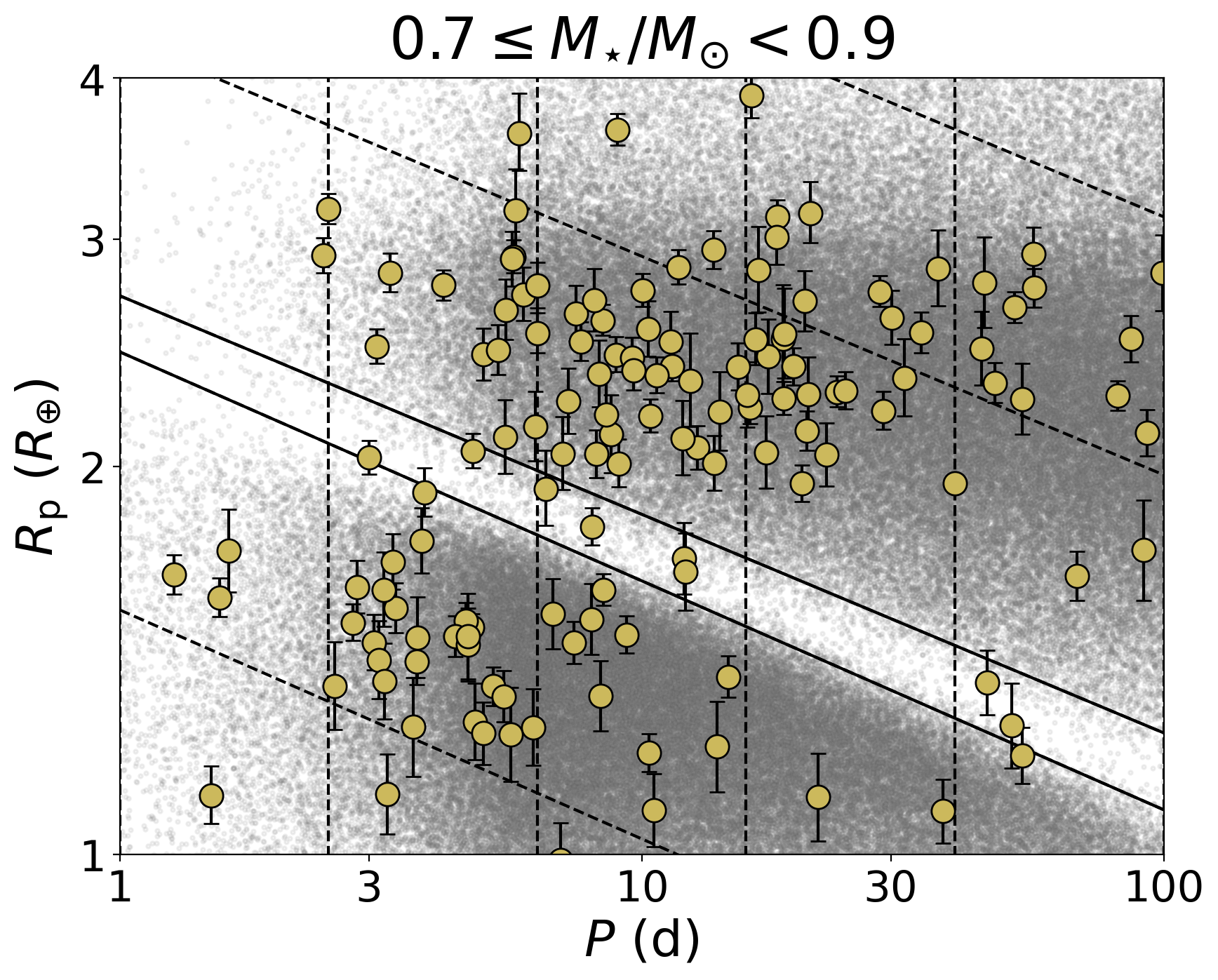} \\
        \includegraphics[width=\columnwidth]{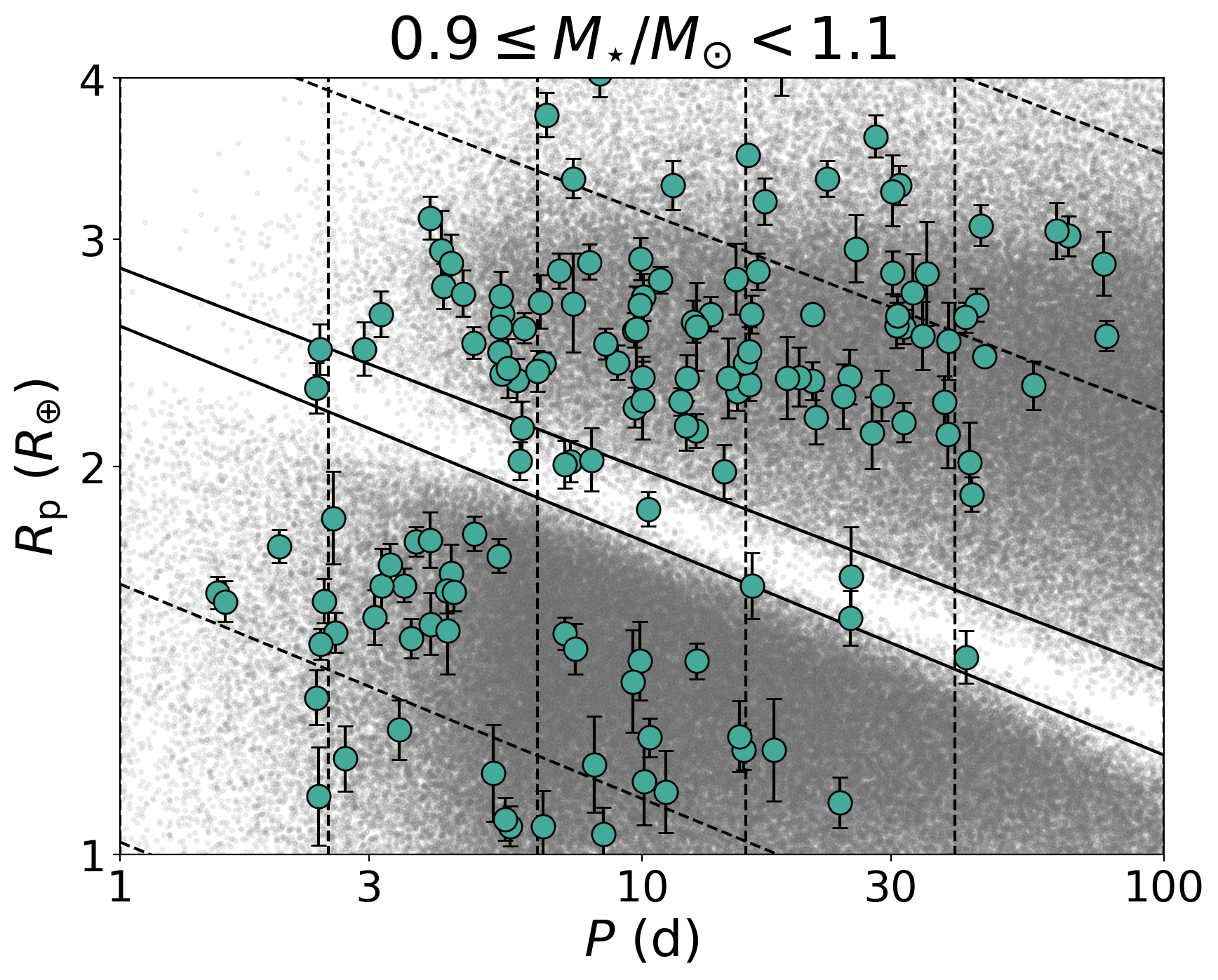} & \includegraphics[width=\columnwidth]{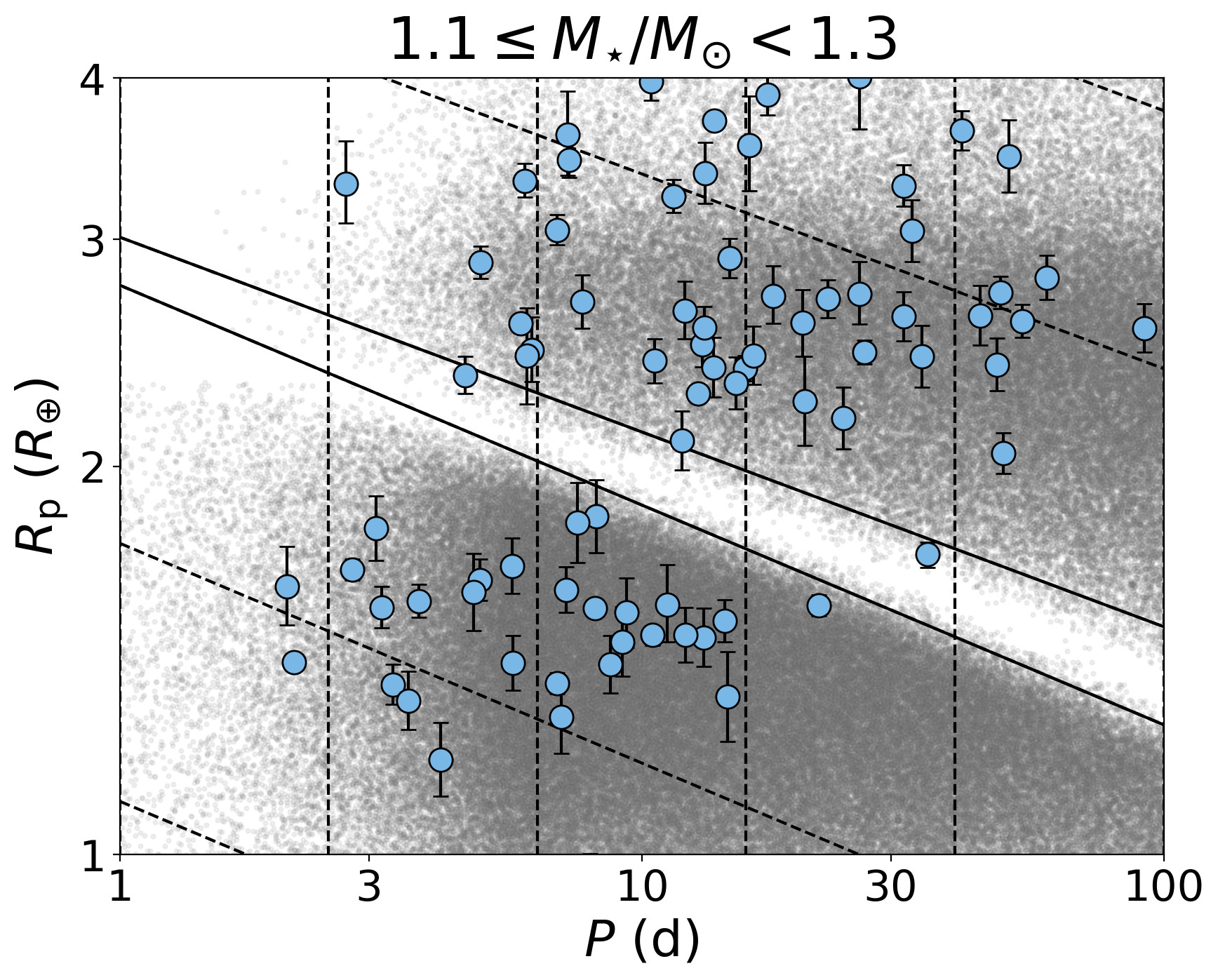}
    \end{tabular}
    \caption{Radius-orbital period plot of planets for different host star masses. Grey dots indicate planets simulated with the XUV photoevaporation models of \citet{rogers2021photoevaporation}, and coloured circles indicate the observed planets. The black solid lines show the boundaries of the model's radius valley, as defined by the parameters in Table~\ref{tab:model_slopes}. The black dotted lines show the cells defined in Section~\ref{subsect:model_obs_compare}.}
    \label{fig:RP_cell_model_obs}
\end{figure*}

We plot $10^6$ modelled planets together with the observed sample in Fig.~\ref{fig:RP_cell_model_obs}. Building on methods discussed in \citet{rogers2021photoevaporation} and \citet{petigura2022california}, we locate the boundaries of the radius valley region as the area enclosing 1\% of the modelled planets, while maximising the valley area. Table~\ref{tab:model_slopes} lists the upper and lower boundaries of the radius valley in the photoevaporation model for each stellar mass range. We compute the area of the radius valley of the photoevaporation model in the $R_{\text{p}}-P$ diagram for different stellar mass bins, and report this in Table~\ref{tab:model_slopes}. We see that the area of the radius valley increases with increasing stellar mass, meaning that higher-mass stars produce a wider radius valley. To first order, the width of the valley is controlled by the difference in size between a stripped core of a given mass and the same core if it were to host a small hydrogen-dominated atmospheric mass fraction $\sim 0.1\%$. The cause of the increase in valley width is that planets orbiting higher-mass stars experience a higher equilibrium temperature at a given orbital separation. Under these circumstances and all else held fixed, the transit radius of a given sub-Neptune is more inflated when compared to those around smaller-mass stars. On a population level, this increases the difference in size between a stripped core and sub-Neptune, thus widening the valley.

The model's widening valley with increasing stellar mass may seem to qualitatively match the observations, where a narrower valley could appear shallower due to the uncertainty of the observed planet radius. However, further steps are required to robustly compare our observations to models, as observational biases exist that are absent in the models, meaning the observed planet population has a different distribution in radius-period space compared to the models. Therefore, we perform a more rigorous comparison in Section~\ref{subsect:model_obs_compare}.

\begin{table}
    \centering
    \caption{Slopes and intercepts of lower and upper radius valley boundaries of the photoevaporation model used in this work, given by $\log_{10}{\left(R_{\text{p}}/R_{\earth} \right)} = m\log_{10}{\left(P/\text{d} \right)} + c$. The area of the radius valley, incorporating 1\% of the modelled planets, is also reported.}
    \begin{tabular}{cccccc}
        \hline
         $M_{\star}/M_{\sun}$ & $m_{\text{lower}}$ & $c_{\text{lower}}$ & $m_{\text{upper}}$ & $c_{\text{upper}}$ & Valley area \\
         \hline
         $0.5-0.7$ & $-0.1846$ & $0.3528$ & $-0.1846$ & $0.4021$ & 0.0988 \\
         $0.7-0.9$ & $-0.1775$ & $0.3898$ & $-0.1694$ & $0.4333$ & 0.1031 \\
         $0.9-1.1$ & $-0.1663$ & $0.4099$ & $-0.1559$ & $0.4548$ & 0.1106 \\
         $1.1-1.3$ & $-0.1704$ & $0.4415$ & $-0.1510$ & $0.4787$ & 0.1133 \\
         \hline
    \end{tabular}
    \label{tab:model_slopes}
\end{table}

\subsection{Comparison with theoretical models of atmospheric mass loss} \label{subsect:model_obs_compare}

As discussed in Section \ref{subsect:results_model}, we wish to compare \textit{Kepler} data with a realisation of the photoevaporation model that fits the observations as well as reasonably possible. In Section \ref{subsect:results_model}, we constructed models that minimised the number of observed planets inside the valley, thereby matching the approximate valley slope. However, this did not precisely match the rest of the distribution across the entire range in stellar mass, such as the specific occurrence of super-Earths and sub-Neptunes in various regions of parameter space. In this Section, we introduce a methodology to fit the planet distribution, which avoids the extreme computational costs of \citet{Rogers2021Unveiling}. Our goal is to produce a model that is, in fact, \textit{perfectly} matched to the observations everywhere except inside the valley. This perfectly fit model is then used to predict the number of planets inside the radius valley for each stellar mass bin, accounting for measurement uncertainty, which inevitably scatters some planets into the valley. In this way, we investigate whether the photoevaporation framework is able to match the observed radius valley trend with stellar mass, as a consequence of the model producing a narrower radius valley for lower-mass stars and observational uncertainty.
If the model still cannot reproduce the observed increase of planets inside the valley at lower stellar masses, then the inconsistency is robust. It would imply that photoevaporation models -- even after exhausting the free parameters available within these models -- are insufficient in their current form to explain the trend of the observed radius valley with $M_\star$.

For each stellar mass bin, we divide the radius-period plot into multiple cells, parallel with the valley, as shown in Fig.~\ref{fig:RP_cell_model_obs}. We construct the cells such that the valley is contained within its own cell for each period bin, with boundaries defined from Table \ref{tab:model_slopes}. The goal is to find a numerical weight for each cell outside of the valley such that the number of planets in the model is weighted to exactly match that of the observations. As discussed, this perfectly weighted model is then used to predict the number of planets inside the radius valley for each stellar mass bin, due to measurement uncertainties in planetary radii. To begin, we account for measurement uncertainty in planetary radius by redrawing each modelled planet from $\mathcal{N}\left(R_{\text{p, model}}, \sigma_{R_\text{p,model}} \right)$ where $\sigma_{R_\text{p,model}}$ is a random selection in the set of $\sigma_{R_\text{p,obs}}$ of planets within the same period range. Afterwards, for each cell, we calculate the number of planets that have shifted to another cell as a result of the new draw. We represent this in matrix form:
\begin{equation}
    \mathbf{F} = \begin{bmatrix}
        f_{0,0} & f_{0,1} & \cdots & f_{0,v} & \cdots & f_{0,n} \\
        f_{1,0} & f_{1,1} & \cdots & f_{1,v} & \cdots & f_{1,n} \\
        \vdots & \vdots & \ddots & \vdots & \ddots & \vdots \\ 
        f_{v,0} & f_{v,1} & \cdots & f_{v,v} & \cdots & f_{v,n} \\
        \vdots & \vdots & \ddots & \vdots & \ddots & \vdots \\
        f_{n,0} & f_{n,1} & \cdots & f_{n,v} & \cdots & f_{n,n} \\
        \end{bmatrix}
\end{equation}
where $f_{i,j}$ represents the number of planets that were in cell $i$ originally but are now in cell $j$ after redrawing. Here, $i$, $j \in \{0,1,..,v,..,n\}$ and $v$ represents the valley cell index. At this stage, we approximate to zeroth order that the model's radius valley is completely empty, and remove all elements from $\mathbf{F}$ that associate with the valley cell index $v$, e.g.  $f_{i,v}$ and $f_{v,j}$ being the cells for which planets have shifted into, or out of the valley. Since the measured uncertainties in planetary orbital periods are very small, we calculate $\mathbf{F}$ independently for each period bin since planets do not shift across horizontal cells as a result of the new draws. We can then calculate the weights for each cell 
\begin{equation}
    \bm{w} = \begin{bmatrix}
        w_0 & w_1 & ... & w_v & ... & w_n
        \end{bmatrix}
\end{equation}
by solving the matrix equation
\begin{equation}
    \bm{n_{\text{obs}}} = \mathbf{F} \cdot \bm{w}
    \label{eq:matrix_nobs}
\end{equation}
where 
\begin{equation}
    \bm{n_{\text{obs}}} = \begin{bmatrix} 
                                n_{\text{obs},0} & n_{\text{obs},1} & n_{\text{obs},2} & \cdots & n_{\text{obs},n}
                                \end{bmatrix}
\end{equation}
is the array of the number of observed planets in each cell. This is achieved by finding $\mathbf{F}^{-1}$ (the inverse of $\mathbf{F}$), i.e.,
\begin{equation}
    \mathbf{F}^{-1} = \frac{1}{\det \left(\mathbf{F} \right)} \cdot  \mathrm{adj}\left(\mathbf{F} \right)
\end{equation}
which is numerically trivial since $\mathbf{F}$ is a square matrix with dominant diagonal elements. To prevent a singular matrix, we assign a value of 1 for the diagonal terms $f_{i,i}$ if cell $i$ has no planets inside. Our only requirement for the weight inside the valley, $w_v$, is that the weights vary smoothly across the valley for increasing planet size. As such, we determine $w_v$ by taking the average of the two adjacent weights, $w_{v-1}$ and $w_{v+1}$, i.e.
\begin{equation}
    w_v = \frac{w_{v+1} + w_{v-1}}{2}.
\end{equation}
Our perfectly weighted model is now used to calculate the expected number of planets in the model's radius valley with
\begin{equation}
    n_{\text{model,valley}} = \bm {f_{\text{valley}}} \cdot \bm{w}
\end{equation}
where 
\begin{equation}
    \bm {f_{\text{valley}}} = \begin{bmatrix}
                        f_{0,\text{v}} & f_{1,\text{v}} & f_{2,\text{v}} & \cdots & f_{n,\text{v}}
                        \end{bmatrix}
\end{equation}
is the vector containing the number of planets that have shifted from different cells into the radius valley from the new draw. In essence, $n_{\text{model,valley}}$ represents the number of planets the photoevaporation model would predict to exist inside the valley, given that it reproduces the observations everywhere else. 

\begin{figure}
    \centering
    \includegraphics[width=\columnwidth]{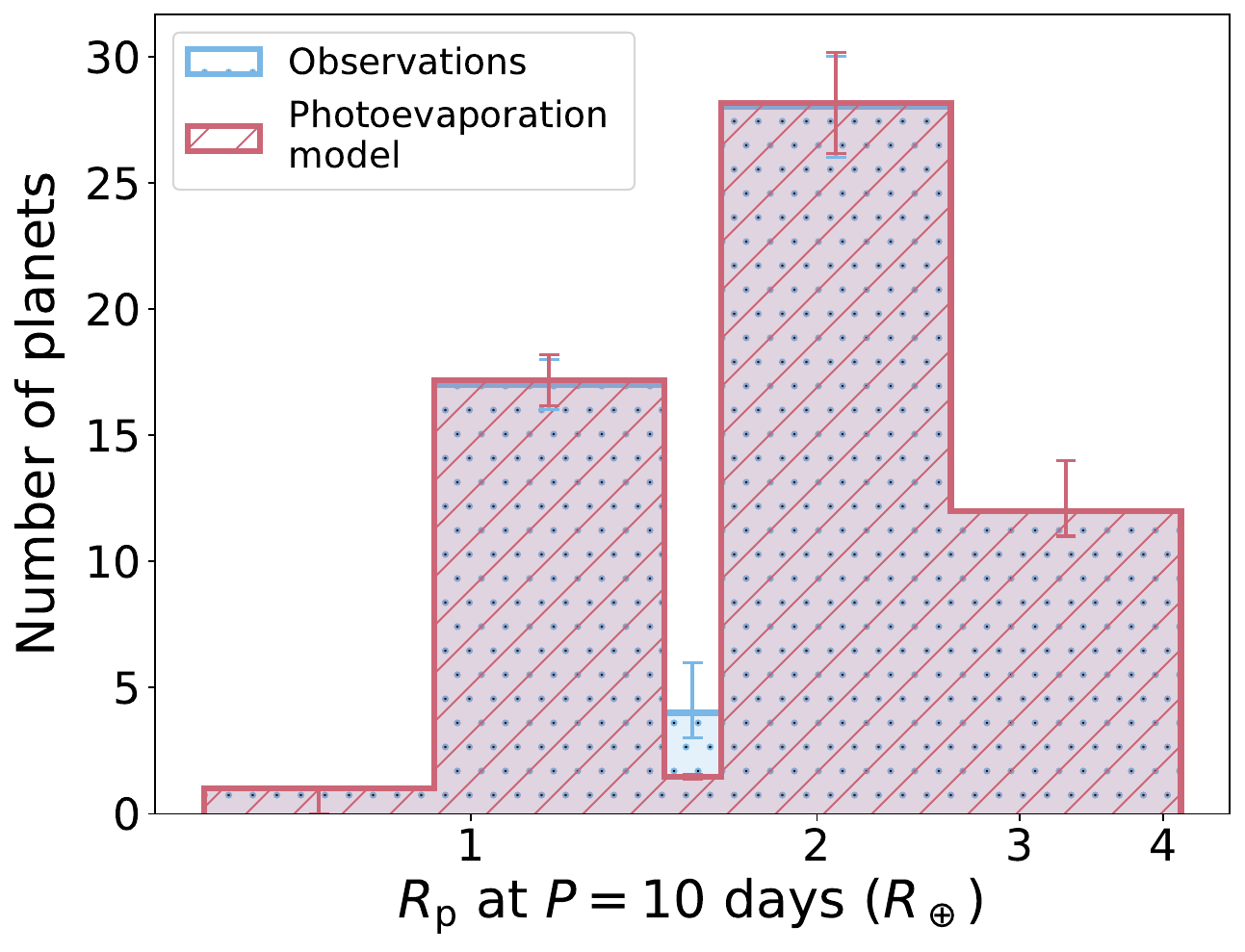}
    \caption{Histogram of $R_{\text{p}}$, adjusted to an equivalent radii at $P = 10$~d, according to the slope of the model's radius valley as listed in Table~\ref{tab:model_slopes}, after weighting the model according to the method discussed in Section~\ref{subsect:model_obs_compare}. Here, the number of planets match everywhere except inside the valley.}
    \label{fig:histcell_pe_obs_0507}
\end{figure}

We perform a bootstrap of 1000 samples for each stellar mass bin, each time randomly redrawing a new set of $R_{\text{p,obs}}$ from $\mathcal{N}\left(R_{\text{p,obs}}, \sigma_{R_{\text{p,obs}}} \right)$, and solving equation~\ref{eq:matrix_nobs} to get a new $n_{\text{model,valley}}$, to obtain the median and $\pm 1 \sigma$ uncertainties of $n_{\text{model,valley}}$ and $n_{\text{obs,valley}}$. To investigate the effect of bin sizes in this analysis, we tested this method with multiple bin sizes in period and radius. We repeat the above-mentioned steps for different equally-spaced logarithmic bins in periods between $1$ and $100$~d, and varying bin sizes in radius. We find that the resulting values for $n_{\text{model,valley}}$ do not change substantially, and all values are within $1 \sigma$ of each other. We therefore use 5 equally-spaced bins in $\log_{10}{P}$ between 1 and 100 d, and a bin width of 0.2 in $\log_{10}{R_{\text{p}}}$, for reporting the final result. Fig.~\ref{fig:histcell_pe_obs_0507} shows the histogram comparing the number of planets in the lowest stellar mass bin ($0.5 \leq M_{\star}/M_{\sun} < 0.7$) between the observations and the photoevaporation model after weighting; here we see that the two match everywhere outside the valley, however there is an excess of planets inside the valley for the observations.

\begin{figure}
    \centering
    \includegraphics[width=\columnwidth]{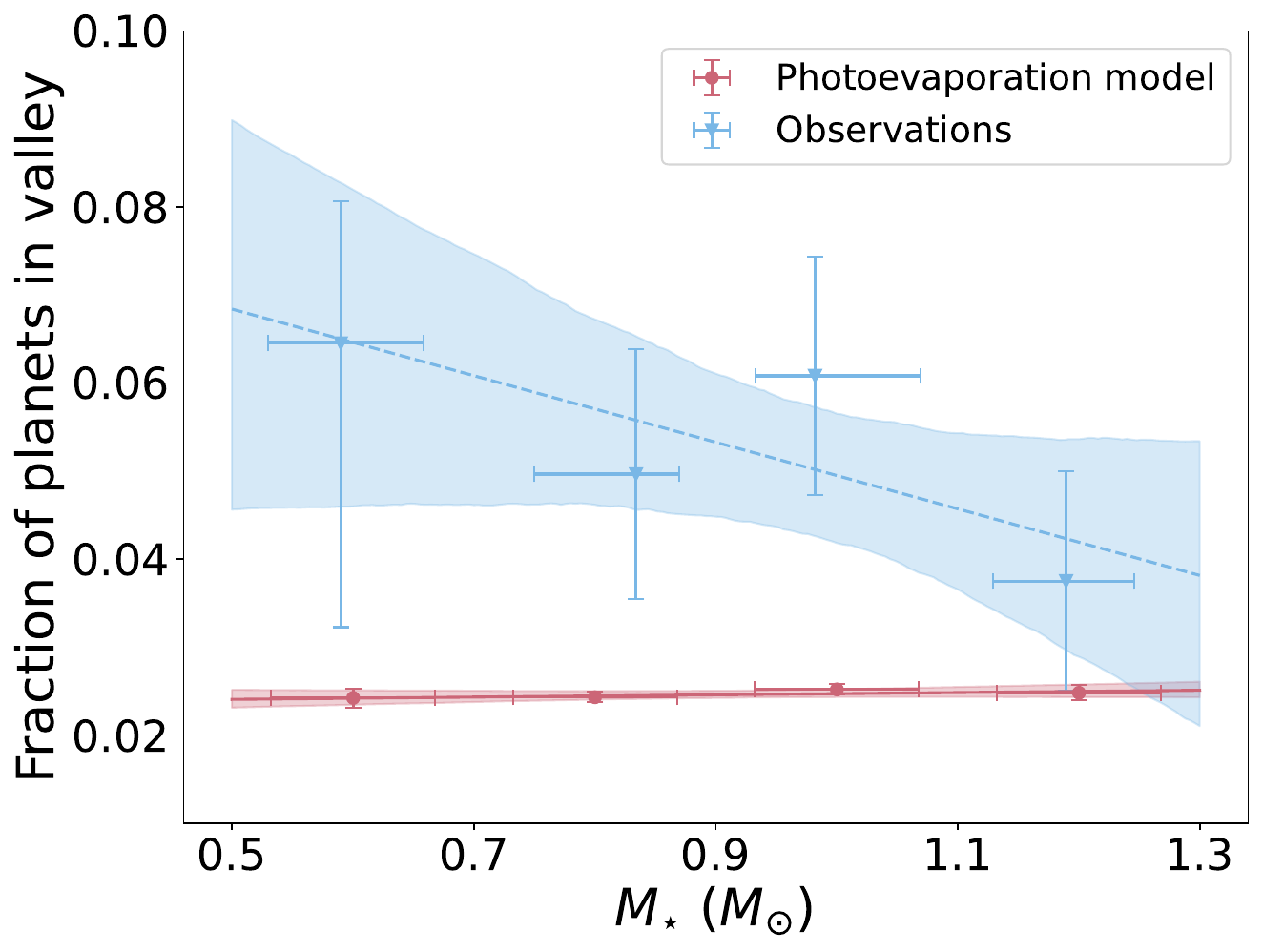}
    \caption{Fraction of planets inside the radius valley as a function of stellar mass, after eliminating effects of initial conditions in the model. The lines show the linear best fit to the data points. We find a slope and intercept of $-0.038^{+0.044}_{-0.045}$ and $0.087^{+0.042}_{-0.045}$ respectively for the observations, and $0.001^{+0.002}_{-0.002}$ and $0.023^{+0.002}_{-0.002}$ for the photoevaporation model.}
    \label{fig:fracgap}
\end{figure}

Fig.~\ref{fig:fracgap} shows the fraction of planets inside the radius valley as a function of stellar mass, comparing the photoevaporation model with observations. The results show that for the observations, this fraction largely decreases with increasing stellar mass, with a best-fitting slope of $m = -0.038^{+0.044}_{-0.045}$, whereas in the theoretical photoevaporation scenario, the fraction stays approximately constant, with $m = 0.001^{+0.002}_{-0.002}$. Since the area of the radius valley also decreases with decreasing stellar mass (see Table~\ref{tab:model_slopes}), the stellar mass dependence of the fraction is less significant compared to the $E$ metrics computed in Section~\ref{subsect:results_shallow} (see Fig.~\ref{fig:E_bymstar}), which is independent of the radius valley area. However, even after taking this effect into account, we continue to see the decrease in fraction with increasing stellar mass, meaning that for lower-mass stars, there are more planets inside a narrower valley for the observations. The measured fractions between the model and observations are more than $1\sigma$ apart for the 3 lowest stellar mass data points, with the fraction of planets inside the radius valley much higher for observations compared to the photoevaporation case. We therefore conclude that current photoevaporation models under-predict the number of planets observed inside the radius valley for low-mass stars. This suggests that these models need to be modified or that further physical mechanisms need to be considered to explain the characteristics of small planets orbiting low-mass stars.

\section{Causes for discrepancy between observation and model} \label{sect:discussion}
\begin{figure*}
    \centering
    \begin{tabular}{ccc}
        \includegraphics[width=0.31\linewidth]{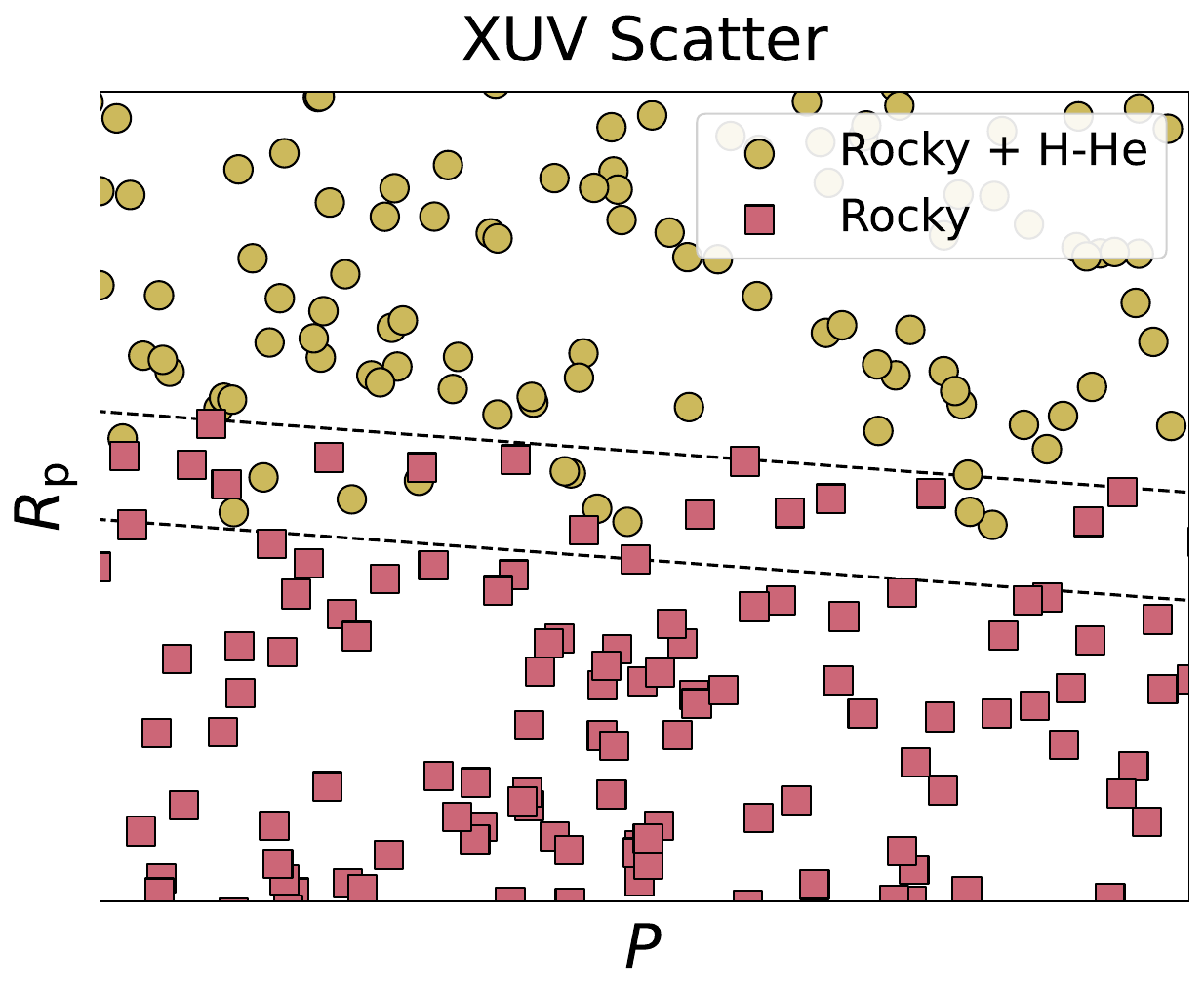} & \includegraphics[width=0.31\linewidth]{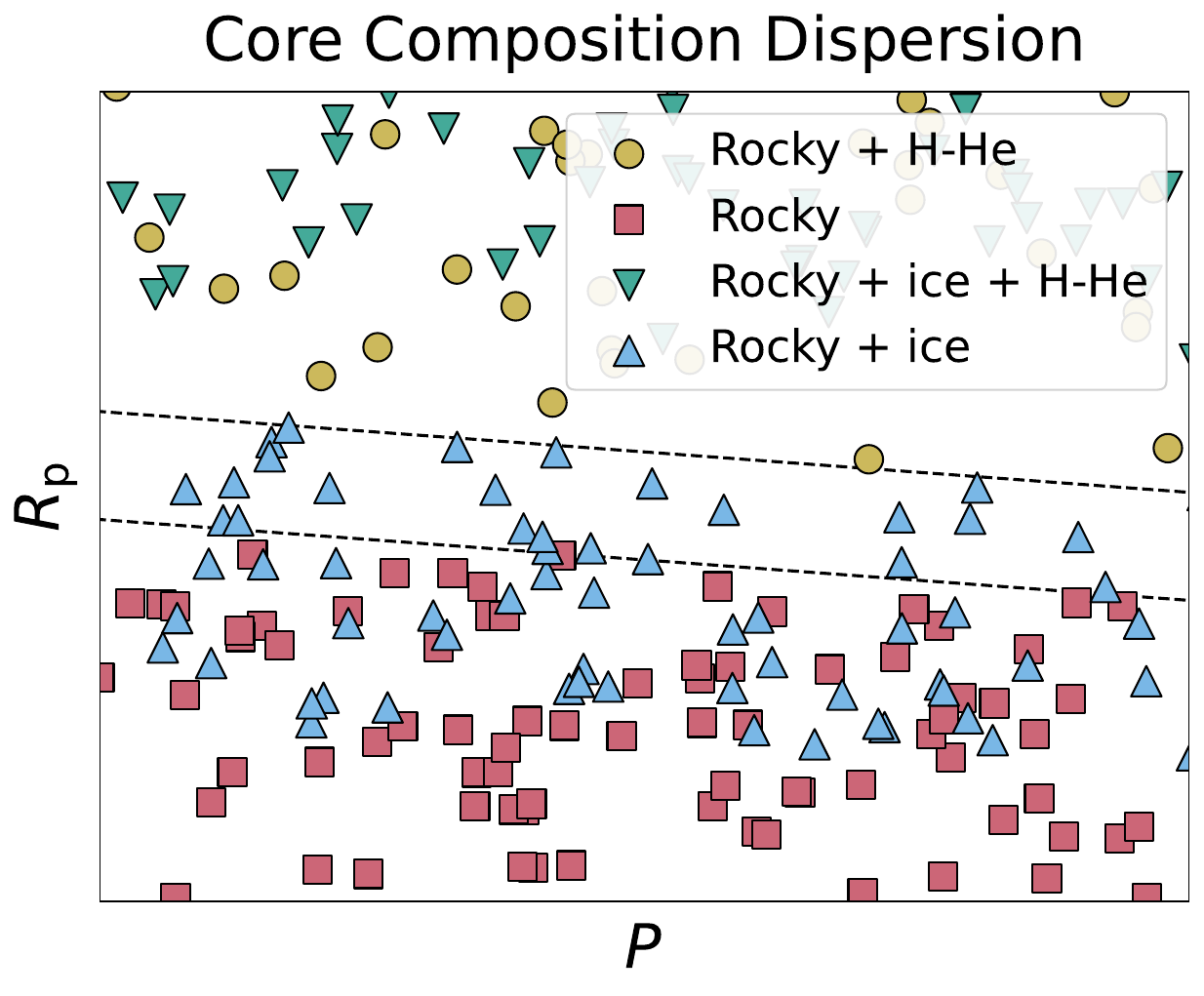} & \includegraphics[width=0.31\linewidth]{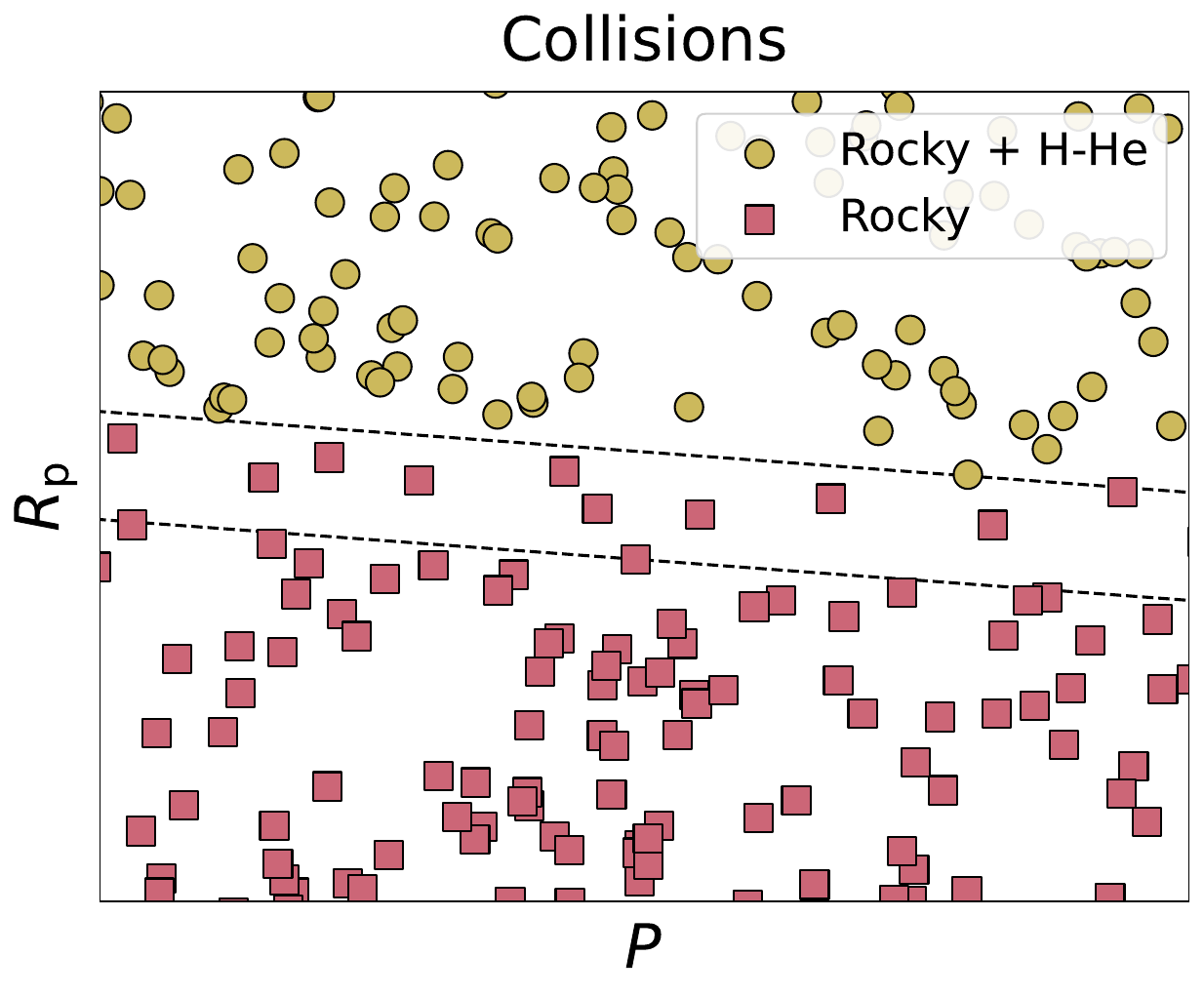}
    \end{tabular}
    \caption{Hypothetical $R_{\text{p}}$-$P$ plots of XUV scatter (left panel), dispersion in planetary core compositions (middle panel), and collisions (right panel). Planets of different types are represented by different colours and shapes. The radius valley is bounded by the black dotted lines.}
    \label{fig:discussion_cartoon}
\end{figure*}

In Section~\ref{sect:results}, we noted the increased shallowness of the radius valley for low-mass stars compared to their high-mass counterparts, as well as the discrepancy in the emptiness of the valley between theoretical models and observations. Given the overall success of photoevaporation models in explaining the observed properties of close-in small planets, we investigate how these models can be extended to reconcile them with observations. In particular, we suggest the following possible explanations for the observed discrepancy: X-ray and UV (XUV) scattering (Section~\ref{subsect:xuv_scatter}), dispersion in planetary core composition (Section~\ref{subsect:icy}), and planetary collisions (Section~\ref{subsect:collisions}). Hypothetical plots of how each scenario may present in the radius-period plot are shown in Fig.~\ref{fig:discussion_cartoon}.

\subsection{Scattering in the stars' XUV output} \label{subsect:xuv_scatter}
A star's high-energy output is known to be tightly correlated with its rotation period \citep[e.g.][]{wright2011stellar,johnstone2021active}. It is also known that at a given age and stellar mass, there is a spread in rotation periods, where this spread in rotation periods increases at younger stellar ages \citep[e.g.][]{reinhold2020stellar}. Therefore, empirically, it is anticipated that stars with identical masses could have different histories of their XUV output. Studies \citep[e.g.][]{tu2015extreme,kubyshkina2019kepler} have shown that plausible different histories of a star's high energy outputs could lead to different planetary evolutionary scenarios, where planets with identical initial properties are able to retain or lose an atmosphere. The position of the radius valley directly depends on the XUV exposure ($\mathcal{X}_{\rm xuv}$, i.e. the total lifetime integrated XUV flux incident upon the planet, \citealt{owen2017evaporation}). Thus, as the XUV exposure will vary from star-to-star with the same stellar mass, the radius valley will become blurred. Following \citet{owen2017evaporation} with a mass-loss efficiency scaling of $\eta \propto v_{\rm esc}^{-2}$, with $v_{\rm esc}$ the planet's escape velocity, if a given star's XUV exposure is a fraction $f_{\mathcal{X}}$ of a typical star with the same stellar mass, then the position of the radius valley should scale as:
\begin{equation}
    R_\text{p,valley} \propto f_{\mathcal{X}}^{0.12}.
\end{equation}
This implies a factor of 5 spread in the XUV histories is enough to move the position of the radius valley up or down for an individual star by $\pm10$\%. This is possible as \citet{tu2015extreme} has shown that for planets with the same initial planetary and atmospheric masses, the planet retains 45\% of its initial atmosphere at 5~Gyr when orbiting a slowly rotating star with minimal XUV output, whereas the planet would lose all of its atmosphere at $\sim 100$~Myr if it orbits a fast-rotating star. Thus, variation in a star's XUV history will fill the radius valley; planets around stars with a higher-than-average XUV output will have a radius valley at larger planetary radii. Therefore, more massive cores can be stripped at a given orbital period, polluting the radius valley with rocky planets. However, stars with a lower-than-average XUV output will allow lower mass planets to retain their atmospheres at a given orbital period, polluting the radius valley with planets with a H-He dominated atmosphere and thus appear lower density than a planet with an Earth-like bulk density. Therefore, if XUV scatter is the cause of the excess of planets in the radius valley, planet's in the radius valley should have a range of densities covering Earth-like bulk densities to more volatile-rich densities. 

There are two reasons why this should lead to an increasing pollution factor with decreasing stellar mass. Firstly, as discussed in Section~\ref{subsect:results_model}, the width of the radius valley decreases with decreasing stellar mass. Thus, for a fixed spread in XUV histories, this will naturally pollute a narrower valley more, as any planet is more likely to be scattered into the valley. Secondly, there is some empirical evidence that the scatter in possible XUV histories is larger around lower mass stars, as evidenced by the larger spread in rotation rates measured at a given age \citep[e.g.][]{newton2016rotation,reinhold2020stellar}. Photoevaporation models could be constructed to take XUV scattering into account in order to investigate this effect further; however, since the rotation evolution of M-dwarfs remains poorly constrained, we leave such work for future studies.

\subsection{Dispersion in planetary core composition} \label{subsect:icy}
As in the case of scatter in a planet's XUV history, the bulk density of the planet's core also causes variation in the position of the radius valley. A planet with a lower-density core, but the same mass, has a weaker gravitational potential. Thus, this planet has its atmosphere more easily removed than a planet with a denser core. As such, the radius valley, at a fixed period, appears at a larger radius for lower-density cores. As discussed by \citet{owen2017evaporation, jin2018compositional, mordasini2020planetary}, any spread in core-composition will result in a shallower radius valley (Fig. 8 in \citealt{owen2017evaporation}). Since the position of the radius valley appears to be in agreement with predictions for an Earth-like bulk density, but with an excess of planets in the valley, if a spread in core composition caused this, this would imply that the distribution of core densities has a larger tail of lower bulk density cores around lower mass stars. 

The ice line of stars is the distance from the star beyond which the temperature is low enough for volatiles to condense. Hence, the ice line is closer to the star for lower-mass stars; planets with a larger ice fraction in their cores are more likely to form closer to the star, within the observed orbital period range. Therefore, if there was an increasing population of planets with lower bulk density cores around lower mass stars, this could also explain the excess pollution of the radius valley that we observe \citep[e.g.][]{venturini2020nature}. For planets with higher equilibrium temperatures ($\gtrapprox 400\text{K}$) around low-mass stars, simulations have shown that water in planets exists wholly in the form of steam rather than in liquid or solid phases, and such water mixes with the planetary envelope \citep[e.g.][]{mousis2020irradiated, venturini2024fading}, hence such steam worlds may populate the radius valley for low-mass stars.

\subsection{Collisions} \label{subsect:collisions}
Collisions among planetary bodies have previously been proposed to be an integral part of Super-Earth and Sub-Neptune formation \citep[e.g.][]{inamdar2015formation,izidoro2017breaking}. Such collisions among planets in the super-Earth and Sub-Neptune populations can also lead to a more filled-in radius valley. For example, when two sub-Neptunes collide, this giant impact imparts so much kinetic energy that is subsequently converted into heat that the hydrogen envelope is entirely lost while their cores merge \citep[e.g.][]{liu2015giant,biersteker2019atmospheric,matsumoto2021size,izidoro2022exoplanet}. This, therefore, leads to a more massive but smaller planet (due to the lack of the hydrogen envelope) and hence would populate the radius valley. Similarly,  a merger of sufficiently large super-Earths could also start to populate the radius valley from the lower edge. Regardless of whether the collisions appear between super-Earths and sub-Neptunes or within each of these populations, the net result is generally the loss of any hydrogen-envelope, leaving behind mostly rocky-cores. Therefore, if collisions are responsible for populating the radius valley, then the population of exoplanets residing in it is expected to be predominantly rocky in their composition (right panel of Fig.~\ref{fig:discussion_cartoon}). 

For the same planet mass ($M_{\text{p}}$) and semi-major axis (a), planetary scale collisions are indeed expected to be more common around lower-mass stars because the Hill radius,
\begin{equation}
R_{\rm{Hill}}=a\left(\frac{M_{\text{p}}}{3M_{\star}}\right)^{1/3}
\end{equation}
which estimates the scale over which neighbouring planets can significantly perturb each other gravitationally while orbiting their host star ($M_{\star}$), is larger for lower mass stars. Additionally, at fixed planet mass, encounters can result in perturbations approaching the planets' escape velocity \citep[e.g.][]{goldreich2004planet}, around lower-mass stars (with shallower potentials), these planet-planet scatterings result in higher eccentricities. This, therefore, leads to more eccentricity excitation, eventual orbit crossings and collisions around lower mass stars than higher mass ones, assuming the planet population is identical otherwise. In fact, planets with high bulk densities around low-mass stars, which are thought to be products of collision events \citep[e.g.][]{naponiello2023super} have been discovered.

\subsection{Mass measurements of planets point to a mixture of scenarios} \label{sect:mass_measure}
As discussed in Sections~\ref{subsect:xuv_scatter}-\ref{subsect:collisions}, stellar XUV scatter, dispersion in core compositions, and collisions all result in similar views of the radius valley in radius-period space. To disentangle these degenerate views, one could measure the masses of planets inside the radius valley to infer their compositions.

In our sample of 444 planets analysed in this work, 203 orbit around host stars with $0.5 \leq M_{\star}/M_{\sun} < 0.9$, of which only 33 have archival mass measurements, and only 11 of which have mass precision better than 30\%. This is because the sample studied here originates from \textit{Kepler} observations, which are suited for highly precise and homogeneous transit observations, but are less suited for follow-up mass observations due to the relative faintness of \textit{Kepler} host stars. We increase the sample size by turning to the NASA Exoplanet Archive \citep{akeson2013nasa} for the sample of all confirmed exoplanets with mass measurements, within the period range of $1 \leq P/{\text{d}} \leq 100$ and with planetary radii in the range $1 \leq R_{\text{p}}/R_{\earth} \leq 4$. For planets with multiple reported mass measurements, we select the value published most recently. We also utilise the recent measurements from \citet{leleu2023removing}, which are not yet available on the NASA Exoplanet Archive at the time of writing. Although the most recent mass measurements may not have the best precision, we hold the assumption that the instruments and methodology used, and the data available for retrieving planetary masses from observations, should improve over time. We then only retain planets with mass precision better than 30\%. For radius measurements, we acknowledge the importance of homogeneity and precise radius measurements \citep[e.g.][]{ho2023deep}, hence for planets not in the sample of \citet{ho2023deep} or this work, we use the planetary radii and stellar mass from \citet{berger2023gaia} if available, otherwise we use the corresponding planetary radius and stellar mass measurement reported in the NASA Exoplanet Archive for the same mass entry.

We calculate an expected envelope fraction for the planets by following the method in \citet{chen2016evolutionary}, which applies the Modules for Experiments in Stellar Astrophysics (MESA) models \citep{paxton2011modules, paxton2013modules, paxton2015modules} to sub-Neptune atmospheres. This method takes the planetary radius ($R_{\text{p}}$), planetary mass ($M_{\text{p}}$), planetary incident bolometric flux ($F_{\text{p}}$), and age as inputs. We use $F_{\text{p}}$ from \citet{berger2023gaia} where available, otherwise, we calculate $F_{\text{p}}$ from the stellar radius and effective temperature, and the orbital semi-major axis. We use the stellar ages from \citet{berger2023gaia} if available, else we find the value with the lowest percentage uncertainty from the NASA Exoplanet Archive. We compute $R_{\text{core}}$ using the mass-radius relationship from \citet{fortney2007planetary} for a given $M_{\text{p}}$, assuming an Earth-like rocky/iron core with an iron fraction of 1/3, and that all of the planet's mass reside in its core. We then solve for the $f_{\text{env}}$ required to produce a planet with the given mass and radius using Brent's method of root finding. 
Some planets have mean densities consistent with Earth-like, atmosphere-free compositions, as is expected from photoevaporation models for planets below the radius valley. Where we cannot find an atmospheric solution, we set a default envelope fraction to $10^{-4}$.

\begin{figure*}
    \centering
    \begin{tabular}{cc}
        \includegraphics[width=\columnwidth]{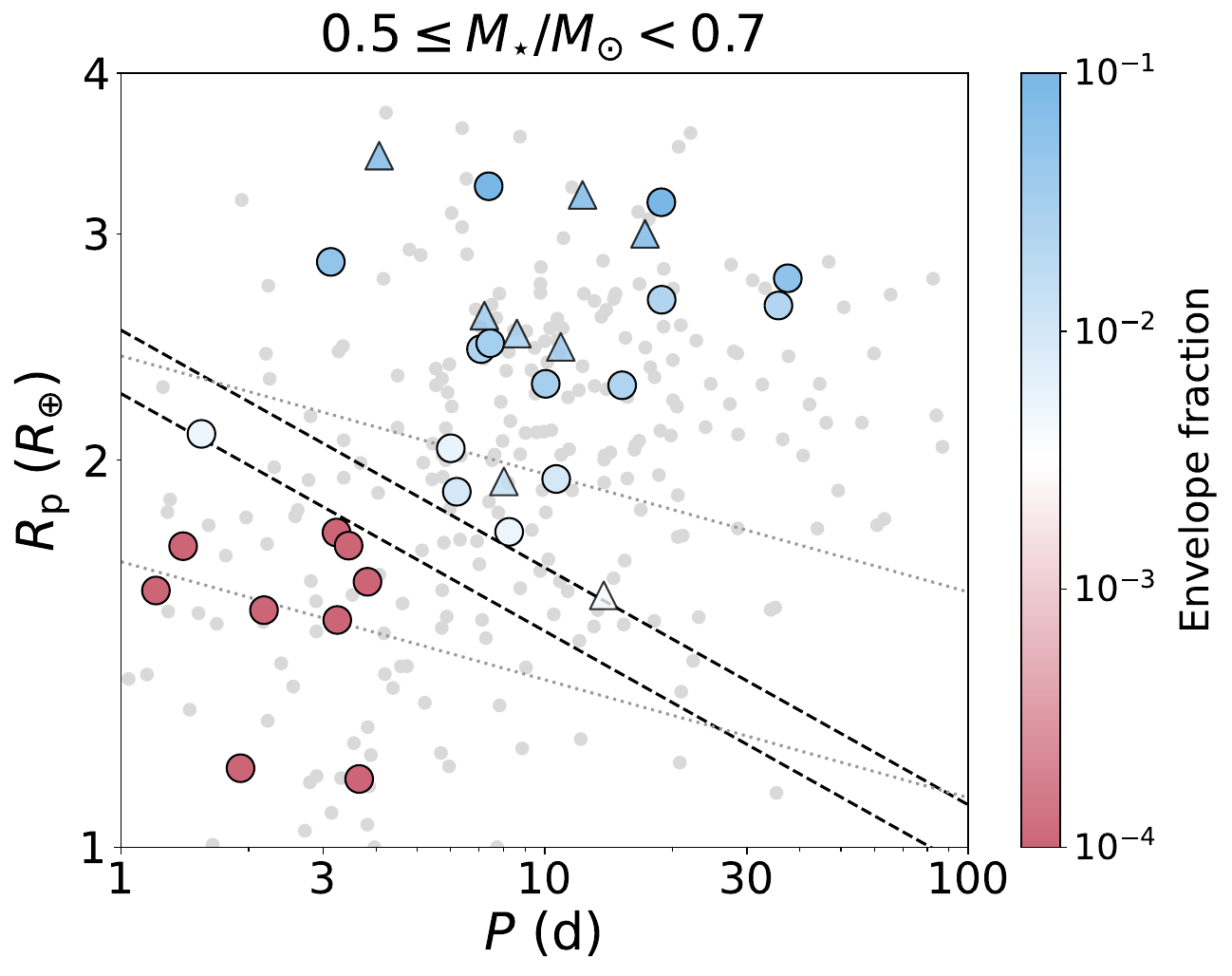} & \includegraphics[width=\columnwidth]{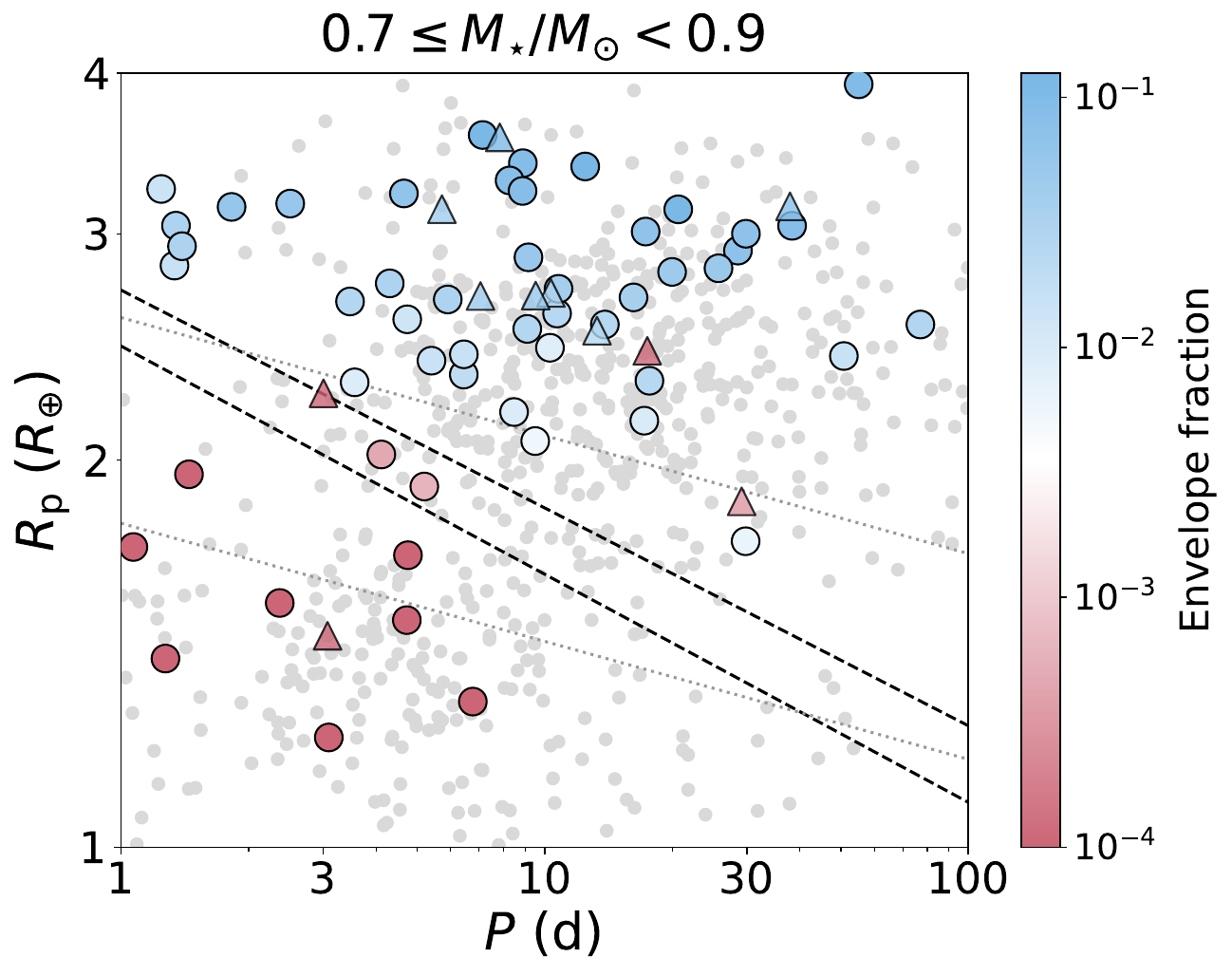}
    \end{tabular}
    \caption{Radius-orbital period plot of all small, close-in planets from the NASA Exoplanet Archive. Coloured points indicate the expected envelope fraction for planets with known mass measurements and with mass precision better than 30\%, and grey points show the remaining planets. The radius valley of the photoevaporation model and the observed valley are bounded by the black dashed lines and grey dotted lines, respectively. Planetary masses are determined from radial velocity (circles) and transit timing variation (triangles) measurements.}
    \label{fig:rp_density}
\end{figure*}

We plot the resulting $f_{\text{env}}$ for planets with mass measurements in a $R_{\text{p}}-P$ plot in Fig.~\ref{fig:rp_density}. A table listing all the planetary masses and envelope fractions is provided in the Appendix (Table~\ref{tab:h24_planet_masses}). We observe that, in general, planets with bare rocky cores lie towards the lower part of the radius-period plot, while planets with significant atmospheric fractions ($\gtrsim 10^{-2}$) populate the upper part. We notice a few planets with a bare rocky core move across the boundary into the region with an abundance of planets with large envelope fractions, however, the masses of these planets mostly come from transit timing variations (TTV) rather than radial velocity (RV) measurements. 

By visual inspection, our results do not wholly match the case for strong XUV scatter, as we do not observe a significant mixing of bare rocky planets and planets with predicted substantial atmospheres inside the valley as in the left panel of Fig.~\ref{fig:discussion_cartoon}. On the other hand, the presence of planets with predicted small, but non-zero, envelope fractions ($\lesssim 10^{-2}$) inside the valley would be consistent with a low ice fraction in their cores, or planets with water-rich atmospheres. Finally, if we consider the TTV masses to be true, the existence of some high-mass bare rocky planets with radii just above the radius valley suggests these are extremely dense planets which may be by-products of collision events. Therefore, we suggest that a mixture of these scenarios is the probable scenario around low-mass stars. 

However, one cannot draw a firm conclusion for the following reasons. Most importantly, we suffer from small number statistics since most planets do not have precise mass measurements available. Another possible concern is that the calculation of the envelope fraction requires information of the stellar age, which are highly uncertain (with a mean uncertainty of $\pm 90\%$ for $0.5 \leq M_{\star}/M_{\sun} < 0.9$ in our sample). However, for a sample of main sequence stars, this is unlikely to significantly affect our conclusions. Nevertheless, a more comprehensive study is required to enhance our understanding around low-mass stars, such as obtaining more precise mass measurements of small planets around low-mass stars, with a focus towards those inside the predicted radius valley, as well as more precise age estimates of such stars, or thorough modelling of XUV scattering or collision events.

\section{Conclusions} \label{sect:conclusions}
In this work, we refitted the light curves of 72 \textit{Kepler} planets around low-mass stars, using \textit{Kepler} 1-minute short cadence data. We present their revised planetary parameters, and combining with the parameters in \citet{ho2023deep}, we statistically analyse the change in the depth of the radius valley as a function of host star mass. 

We find that the radius valley becomes shallower around lower mass host stars. Although the photoevaporation model also predicts a narrower radius valley for lower mass hosts, this filling-in of the observed valley could not be explained by theoretical models of photoevaporation, even when observation biases and dispersion in initial conditions are taken into account. This suggests that additional mechanisms are involved in shaping the radius valley for planets around low-mass stars.

We suggest three possible theoretical scenarios that could explain the shallower radius valley for low-mass stars: scattering in the stars' XUV output, dispersion in planetary core composition, and collision events. Combining our sample with other planets around low-mass stars on the NASA Exoplanet Archive, we calculate the expected envelope fraction of planets under the assumption of a rocky core. We note that the resulting population distribution with reference to the radius valley location does not favour either scenario exclusively, however, it provides evidence for the presence of planets with a low ice/water fraction, or planetary collisions.

With few planets having precise mass and age measurements, we are unable to draw strong conclusions about the true nature of planets around low-mass stars. We suggest potential future work to understand this picture better, such as obtaining more precise planetary mass or stellar age measurements or a comprehensive modelling of XUV scatter or planetary collision events. The Transit Exoplanet Survey Satellite \citep[TESS, e.g.][]{ricker2015transiting} would yield more planets around bright, nearby stars, which are optimal candidates for radial velocity follow-up observations to obtain precise masses. Combining with the future launch of the PLAnetary Transits and Oscillations of stars (PLATO) mission \citep[e.g.][]{rauer2014plato} which enables asteroseismic observations of stars to obtain precise stellar ages, 
we can deepen our insight into the formation and evolutionary trajectories of small, close-in planets orbiting low-mass stars.

\section*{Acknowledgements}

This paper includes data collected by the Kepler mission and obtained from the MAST data archive at the Space Telescope Science Institute (STScI). Funding for the Kepler mission is provided by the NASA Science Mission Directorate. STScI is operated by the Association of Universities for Research in Astronomy, Inc., under NASA contract NAS 5–26555. CSKH would like to thank the Science and Technology Facilities Council (STFC) for funding support through a PhD studentship. JGR is sponsored by the National Aeronautics and Space Administration (NASA) through a contract with Oak Ridge Associated Universities (ORAU). VVE is supported by UK's Science \& Technology Facilities Council (STFC) through the STFC grants ST/W001136/1 and ST/S000216/1. HES gratefully acknowledges support from NASA under grant number 80NSSC21K0392 issued through the Exoplanet Research Program. JEO is supported by a Royal Society University Research Fellowship. This project has received funding from the European Research Council (ERC) under the European Union’s Horizon 2020 Framework Programme (grant agreement no. 853022, PEVAP). We thank the anonymous reviewer for taking their time to review the paper, and for their valuable comments which helped improve the manuscript.


\section*{Data Availability}
The \textit{Kepler} 1-minute short cadence light curves are available for download on the NASA Mikulski Archive for Space Telescopes (MAST) database\footnote{\url{https://archive.stsci.edu}}. The parameter estimates from HMC posteriors for the new planets analysed in this work are provided in the Appendix (Tables~\ref{tab:h24_a1_star},~\ref{tab:h24_a2_planet}, and~\ref{tab:h24_a3_gp}).
 



\bibliographystyle{mnras}
\bibliography{paper} 




\appendix

\section{Extra Material}

\begin{table*}
\renewcommand*{\arraystretch}{1.5}
    \centering
    \caption{Host star parameters of the additional planets fitted in this work, that are not in the sample of \citet{ho2023deep}. Stellar radius ($R_{\star}$), mass ($M_{\star}$), effective temperature ($T_{\text{eff}}$), metallicity ([Fe/H]), Kepler magnitude (Ksmag), and age values, are taken from \citet{petigura2022california}. The `radius correction factors' (RCFs) are taken from \citet{furlan2017kepler}. Stellar densities ($\rho_{\star}$), and the two quadratic limb darkening parameters ($u_0$ and $u_1$) are obtained from the transit fits in this work. Only the first 10 rows are shown here; the full table is available online in a machine-readable format.}
    \begin{tabular}{ccccccccccc}
        \hline 
KOI & $R_{\star}$ ($R_{\sun}$) & $M_{\star}$ ($M_{\sun}$) & $T_{\text{eff}}$ (K) & [Fe/H] (dex) & Ksmag & Age (Gyr) & RCF & $\rho_{\star}$ (g cm$^{-3}$) & $u_0$ & $u_1$ \\ 
\hline 
156 & $0.72_{-0.02}^{+0.02}$ & $0.73_{-0.03}^{+0.03}$ & $4511_{-110}^{+110}$ & $0.11 \pm 0.09$ & $11.37 \pm 0.02$ & $10.1_{-6.5}^{+6.7}$ & 1.0000 & $2.039 \pm 0.164$ & $0.63 \pm 0.07$ & $0.15 \pm 0.10$ \\ 
222 & $0.71_{-0.02}^{+0.02}$ & $0.70_{-0.03}^{+0.03}$ & $4416_{-70}^{+70}$ & $0.09 \pm 0.09$ & $12.31 \pm 0.02$ & $10.7_{-6.6}^{+7.3}$ & 1.0000 & $2.145 \pm 0.180$ & $0.56 \pm 0.09$ & $0.21 \pm 0.13$ \\ 
247 & $0.58_{-0.02}^{+0.02}$ & $0.57_{-0.01}^{+0.01}$ & $3840_{-70}^{+70}$ & $0.05 \pm 0.09$ & $11.12 \pm 0.02$ & $11.6_{-6.5}^{+8.1}$ & 1.0000 & $2.933 \pm 0.157$ & $0.34 \pm 0.12$ & $0.20 \pm 0.13$ \\ 
248 & $0.62_{-0.02}^{+0.02}$ & $0.60_{-0.01}^{+0.01}$ & $4011_{-70}^{+70}$ & $-0.16 \pm 0.09$ & $12.38 \pm 0.02$ & $12.8_{-6.2}^{+9.0}$ & 1.0002 & $2.724 \pm 0.184$ & $0.54 \pm 0.13$ & $0.31 \pm 0.17$ \\ 
249 & $0.41_{-0.01}^{+0.01}$ & $0.40_{-0.01}^{+0.01}$ & $3506_{-70}^{+70}$ & $-0.32 \pm 0.09$ & $11.15 \pm 0.03$ & $11.8_{-6.4}^{+8.1}$ & 1.0000 & $5.748 \pm 0.253$ & $0.28 \pm 0.11$ & $0.44 \pm 0.16$ \\ 
250 & $0.60_{-0.02}^{+0.02}$ & $0.59_{-0.02}^{+0.01}$ & $4055_{-70}^{+70}$ & $-0.17 \pm 0.09$ & $12.63 \pm 0.03$ & $11.6_{-6.6}^{+8.2}$ & 1.0031 & $2.946 \pm 0.245$ & $0.40 \pm 0.14$ & $0.20 \pm 0.13$ \\ 
251 & $0.54_{-0.02}^{+0.02}$ & $0.53_{-0.01}^{+0.01}$ & $3788_{-70}^{+70}$ & $-0.03 \pm 0.09$ & $11.68 \pm 0.02$ & $11.6_{-6.6}^{+8.1}$ & 1.0158 & $3.403 \pm 0.161$ & $0.42 \pm 0.09$ & $0.16 \pm 0.11$ \\ 
252 & $0.63_{-0.02}^{+0.02}$ & $0.60_{-0.01}^{+0.01}$ & $3889_{-70}^{+70}$ & $0.10 \pm 0.09$ & $12.55 \pm 0.03$ & $12.3_{-6.3}^{+8.5}$ & 1.0000 & $2.552 \pm 0.143$ & $0.35 \pm 0.10$ & $0.30 \pm 0.15$ \\ 
253 & $0.66_{-0.02}^{+0.02}$ & $0.63_{-0.01}^{+0.01}$ & $4033_{-70}^{+70}$ & $0.38 \pm 0.09$ & $12.29 \pm 0.04$ & $9.4_{-6.4}^{+5.9}$ & 1.0000 & $2.207 \pm 0.118$ & $0.49 \pm 0.11$ & $0.12 \pm 0.09$ \\ 
312 & $1.45_{-0.05}^{+0.05}$ & $1.26_{-0.04}^{+0.03}$ & $6246_{-100}^{+100}$ & $0.07 \pm 0.06$ & $10.52 \pm 0.01$ & $2.5_{-0.2}^{+0.1}$ & 1.0006 & $0.419 \pm 0.031$ & $0.34 \pm 0.15$ & $0.25 \pm 0.14$ \\ 
... & ... & ... & ... & ... & ... & ... & ... & ... & ... & ... \\ 
\hline
    \end{tabular}
    \label{tab:h24_a1_star}
\end{table*}

\begin{landscape}
\begin{table}
\renewcommand*{\arraystretch}{1.5}
    \centering
    \caption{Planetary parameters of the additional planets fitted in this work, that are not in the sample of \citet{ho2023deep}. Orbital period ($P$), transit ephermeris time ($t_0$), planet-to-star radius ratio ($R_{\text{p}}/R_{\star}$), impact parameter ($b$), eccentricity ($e$) and argument of periapsis ($\omega$) are direct results from the transit fitting, while the planetary radius ($R_{\text{p}}$), ratio between the semi-major-axis and the stellar radius ($a/R_{\star}$) and incident bolometric flux ($F_{\text{p}}$) are indirectly calculated. The $R_{\text{p}}/R_{\star}$ and $R_{\text{p}}$ values from \citet{petigura2022california} (P22) are also listed here for comparison. Only the first 10 rows are shown here; the full table is available online in a machine-readable format.}
    {\scriptsize
    \begin{tabular}{ccccccccccccc}
         \hline 
KOI & Kepler name & $P$ (d) & $t_0$ (BJD-2454833) & $R_{\text{p}}/R_{\star}$ & $R_{\text{p}}/R_{\star}$ (P22) & $R_{\text{p}}$ ($R_{\earth}$) & $R_{\text{p}}$ ($R_{\earth}$) (P22) & $b$ & $e$ & $\omega$ ($^{\circ}$) & $a/R_{\star}$ & $F_{\text{p}}$ ($F_{\earth}$) \\ 
\hline 
K00156.01 & Kepler-114 c & $8.041329 \pm 0.000005$ & $143.0401 \pm 0.0005$ & $0.0229 \pm 0.0004$ & N/A & $1.80_{-0.06}^{+0.06}$ & N/A & $0.58 \pm 0.06$ & $0.05 \pm 0.04$ & $18 \pm 104$ & $19.42 \pm 1.78$ & $45.74 \pm 4.33$ \\ 
K00156.02 & Kepler-114 b & $5.188560 \pm 0.000004$ & $145.3632 \pm 0.0005$ & $0.0172 \pm 0.0003$ & $0.0172_{-0.0005}^{+0.0005}$ & $1.35_{-0.05}^{+0.05}$ & $1.37_{-0.06}^{+0.06}$ & $0.51 \pm 0.09$ & $0.05 \pm 0.04$ & $19 \pm 102$ & $14.52 \pm 1.31$ & $81.83 \pm 7.66$ \\ 
K00156.03 & Kepler-114 d & $11.776137 \pm 0.000005$ & $142.7052 \pm 0.0004$ & $0.0364 \pm 0.0005$ & $0.0343_{-0.0008}^{+0.0008}$ & $2.85_{-0.09}^{+0.09}$ & $2.73_{-0.11}^{+0.11}$ & $0.65 \pm 0.04$ & $0.04 \pm 0.03$ & $-8 \pm 102$ & $24.52 \pm 1.81$ & $28.68 \pm 2.23$ \\ 
K00222.01 & Kepler-120 b & $6.312501 \pm 0.000005$ & $132.6551 \pm 0.0006$ & $0.0328 \pm 0.0007$ & $0.0323_{-0.0008}^{+0.0008}$ & $2.54_{-0.09}^{+0.09}$ & $2.53_{-0.10}^{+0.10}$ & $0.39 \pm 0.13$ & $0.05 \pm 0.03$ & $13 \pm 105$ & $16.73 \pm 1.50$ & $56.62 \pm 5.15$ \\ 
K00222.02 & Kepler-120 c & $12.794560 \pm 0.000012$ & $130.7675 \pm 0.0009$ & $0.0267 \pm 0.0007$ & $0.0258_{-0.0009}^{+0.0009}$ & $2.07_{-0.08}^{+0.08}$ & $2.02_{-0.09}^{+0.09}$ & $0.42 \pm 0.13$ & $0.05 \pm 0.04$ & $19 \pm 101$ & $26.97 \pm 2.49$ & $21.77 \pm 2.04$ \\ 
K00247.01 & KOI-247.01 & $13.815046 \pm 0.000014$ & $181.1253 \pm 0.0007$ & $0.0304 \pm 0.0010$ & $0.0294_{-0.0020}^{+0.0020}$ & $1.92_{-0.09}^{+0.09}$ & $1.88_{-0.14}^{+0.14}$ & $0.49 \pm 0.23$ & $0.37 \pm 0.14$ & $87 \pm 52$ & $45.74 \pm 7.46$ & $4.33 \pm 0.71$ \\ 
K00248.03 & Kepler-49 d & $2.576569 \pm 0.000002$ & $172.1278 \pm 0.0006$ & $0.0267 \pm 0.0006$ & $0.0268_{-0.0004}^{+0.0004}$ & $1.81_{-0.07}^{+0.07}$ & $1.81_{-0.06}^{+0.06}$ & $0.59 \pm 0.06$ & $0.05 \pm 0.04$ & $13 \pm 108$ & $9.97 \pm 0.92$ & $108.51 \pm 10.15$ \\ 
K00248.04 & Kepler-49 e & $18.596055 \pm 0.000018$ & $141.2661 \pm 0.0008$ & $0.0288 \pm 0.0011$ & $0.0263_{-0.0008}^{+0.0008}$ & $1.95_{-0.10}^{+0.10}$ & $1.77_{-0.08}^{+0.08}$ & $0.81 \pm 0.03$ & $0.05 \pm 0.04$ & $7 \pm 105$ & $37.04 \pm 3.35$ & $7.86 \pm 0.72$ \\ 
K00249.01 & Kepler-504 A b & $9.549277 \pm 0.000007$ & $175.7568 \pm 0.0005$ & $0.0396 \pm 0.0011$ & $0.0392_{-0.0008}^{+0.0008}$ & $1.77_{-0.07}^{+0.07}$ & $1.77_{-0.07}^{+0.07}$ & $0.39 \pm 0.21$ & $0.38 \pm 0.11$ & $94 \pm 38$ & $45.08 \pm 6.07$ & $3.10 \pm 0.42$ \\ 
K00250.03 & Kepler-26 d & $3.543905 \pm 0.000004$ & $136.2606 \pm 0.0008$ & $0.0184 \pm 0.0006$ & $0.0208_{-0.0004}^{+0.0004}$ & $1.20_{-0.05}^{+0.05}$ & $1.37_{-0.05}^{+0.05}$ & $0.43 \pm 0.13$ & $0.05 \pm 0.04$ & $13 \pm 102$ & $12.67 \pm 1.18$ & $70.16 \pm 6.67$ \\ 
... & ... & ... & ... & ... & ... & ... & ... & ... & ... & ... & ... & ... \\ 
\hline
    \end{tabular}
    }
    \label{tab:h24_a2_planet}
\end{table}
\end{landscape}

\begin{table}
    \centering
    \caption{Transit jitter and Gaussian Process (GP) parameters for the transit fitting of planetary systems performed in this work. Only the first 10 rows are shown here; the full table is available online in a machine-readable format.}
    \begin{tabular}{cccc}
         \hline 
KOI & $\log{\sigma_{\text{lc}}}$ & $\log{\sigma_{\text{gp}}}$ & $\log{\rho_{\text{gp}}}$ \\ 
\hline 
156 & $-7.0732 \pm 0.0011$ & $-8.8706 \pm 0.0167$ & $-3.7099 \pm 0.0404$ \\ 
222 & $-6.4394 \pm 0.0029$ & $-9.4256 \pm 1.5430$ & $-4.4393 \pm 1.7274$ \\ 
247 & $-6.6126 \pm 0.0043$ & $-8.4369 \pm 0.0699$ & $-3.9891 \pm 0.1803$ \\ 
248 & $-6.1755 \pm 0.0014$ & $-7.9431 \pm 0.0193$ & $-4.0146 \pm 0.0470$ \\ 
249 & $-6.5927 \pm 0.0032$ & $-8.6823 \pm 0.0644$ & $-3.4745 \pm 0.1402$ \\ 
250 & $-6.1652 \pm 0.0016$ & $-7.9600 \pm 0.0210$ & $-3.6427 \pm 0.0523$ \\ 
251 & $-6.3174 \pm 0.0020$ & $-8.3393 \pm 0.0431$ & $-3.8431 \pm 0.1113$ \\ 
252 & $-5.9738 \pm 0.0033$ & $-7.6507 \pm 0.0399$ & $-3.8177 \pm 0.0984$ \\ 
253 & $-5.7038 \pm 0.0021$ & $-7.9143 \pm 0.0578$ & $-3.6664 \pm 0.1233$ \\ 
312 & $-8.0250 \pm 0.0061$ & $-9.6177 \pm 0.0736$ & $-3.7928 \pm 0.2148$ \\ 
... & ... & ... & ... \\ 
\hline
    \end{tabular}
    \label{tab:h24_a3_gp}
\end{table}

\begin{landscape}

\begin{table}
\renewcommand*{\arraystretch}{1.5}
    \centering
    \caption{Planetary parameters of 95 planets with $0.5 \leq M_{\star}/M_{\sun} < 0.9$,  with archival planetary mass measurements available. The envelope fraction is calculated using the method described in Section~\ref{sect:mass_measure}. Only the first 10 rows are shown here; the full table is available online in a machine-readable format. Sources: 1. This work, 2. \citet{berger2023gaia}, 3. \citet{dragomir2019tess}, 4. \citet{passegger2024compact}, 5. \citet{dumusque2019hot}, 6. \citet{orellmiquel2023hd}, 7. \citet{barros2023young}, 8. \citet{heidari2022hd}, 9. \citet{rosenthal2021california}, 10. \citet{barros2022hd}, 11. \citet{bonomo2023cold}, 12. \citet{bourrier2022cheops}, 13. \citet{elmufti2023toi}, 14. \citet{diaz2020magellan}, 15. \citet{ellis2021directly}, 16. \citet{vanderburg2015characterizing}, 17. \citet{macdougall2021tess}, 18. \citet{mayo2023hyades}, 19. \citet{lopez2019exoplanet}, 20. \citet{korth2019k2}, 21. \citet{akanamurphy2021another}, 22. \citet{persson2018super}, 23. \citet{mortier2018k2}, 24. \citet{lam2018k2}, 25. \citet{palle2019detection}, 26. \citet{lillobox2020masses}, 27. \citet{piaulet2023evidence}, 28. \citet{ho2023deep}, 29. \citet{hadden2014densities}, 30. \citet{fulton2018california}, 31. \citet{vaneylen2018asteroseismic}, 32. \citet{leleu2023removing}, 33. \citet{vissapragada2020diffuser}, 34. \citet{hadden2017kepler}, 35. \citet{sun2019kepler}, 36. \citet{macdonald2016dynamical}, 37. \citet{weiss2024kepler}, 38. \citet{astudillodefru2020hot}, 39. \citet{west2019ngts}, 40. \citet{otegi2021tess}, 41. \citet{wilson2022pair}, 42. \citet{korth2023toi}, 43. \citet{luque2022density}, 44. \citet{turtelboom2022tess}, 45. \citet{nielsen2020mass}, 46. \citet{lam2023discovery}, 47. \citet{deeg2023toi}, 48. \citet{martioli2023toi}, 49. \citet{desai2024tess}, 50. \citet{espinoza2022transiting}, 51. \citet{leleu2021six}, 52. \citet{mallorquin2023toi}, 53. \citet{naponiello2023super}, 54. \citet{dai2023mini}, 55. \citet{hoyer2021toi}, 56. \citet{suarezmascareno2024tess}, 57. \citet{kunimoto2023toi}, 58. \citet{carleo2020multiplanet}, 59. \citet{osborn2021toi}, 60. \citet{osborne2024toi}, 61. \citet{brinkman2023toi}, 62. \citet{luque2021planetary}, 63. \citet{burt2020toi}, 64. \citet{hawthorn2023toi}, 65. \citet{lillobox2023toi}, 66. \citet{maciejewski2023hot}, 67. \citet{peterson20182}, 68. \citet{polanski2021wolf}.}
    {\scriptsize
    \begin{tabular}{cccccccccccccc}
        \hline 
Planet & $P$ (days) & $R_{\text{p}}$ ($R_{\earth}$) & $R_{\text{p}}$ source & $M_{\text{p}}$ ($M_{\earth}$) & $M_{\text{p}}$ source & RV/TTV mass & $M_{\star}$ ($M_{\sun}$) & $M_{\star}$ source & $F_{\text{p}}$ ($F_{\earth}$) & $S$ source & Age (Gyr) & Age source & Envelope fraction \\ 
\hline 
GJ 143 b & $35.613445_{-0.000043}^{+0.000043}$ & $2.64_{-0.42}^{+0.41}$ & 2 & $22.70_{-1.90}^{+2.20}$ & 3 & RV & $0.69_{-0.04}^{+0.04}$ & 2 & $5.63_{-0.48}^{+0.31}$ & 2 & $15.08_{-9.99}^{+8.77}$ & 2 & 0.0233 \\ 
GJ 9827 b & $1.208912_{-0.000131}^{+0.000131}$ & $1.58_{-1.37}^{+1.37}$ & 2 & $4.28_{-0.33}^{+0.35}$ & 4 & RV & $0.60_{-0.02}^{+0.02}$ & 2 & $269.58_{-12.85}^{+13.16}$ & 2 & $13.25_{-9.49}^{+9.74}$ & 2 & 0.0001 \\ 
GJ 9827 c & $3.648103_{-0.000010}^{+0.000013}$ & $1.13_{-0.05}^{+0.07}$ & 4 & $1.86_{-0.39}^{+0.37}$ & 4 & RV & $0.60_{-0.02}^{+0.02}$ & 2 & $60.35_{-4.11}^{+4.20}$ & 2,4 & $13.25_{-9.49}^{+9.74}$ & 2 & 0.0001 \\ 
GJ 9827 d & $6.201812_{-0.000009}^{+0.000009}$ & $1.89_{-0.14}^{+0.16}$ & 4 & $3.02_{-0.57}^{+0.58}$ & 4 & RV & $0.60_{-0.02}^{+0.02}$ & 2 & $29.70_{-1.96}^{+2.01}$ & 2,4 & $13.25_{-9.49}^{+9.74}$ & 2 & 0.0095 \\ 
HD 15337 b & $4.755974_{-0.000018}^{+0.000018}$ & $1.69_{-0.69}^{+0.69}$ & 2 & $7.20_{-0.81}^{+0.81}$ & 5 & RV & $0.83_{-0.06}^{+0.04}$ & 2 & $161.22_{-8.86}^{+7.06}$ & 2 & $7.90_{-5.93}^{+9.24}$ & 2 & 0.0001 \\ 
HD 15337 c & $17.180721_{-0.000046}^{+0.000046}$ & $2.15_{-1.60}^{+1.60}$ & 2 & $8.79_{-1.68}^{+1.68}$ & 5 & RV & $0.83_{-0.06}^{+0.04}$ & 2 & $29.09_{-1.60}^{+1.27}$ & 2 & $7.90_{-5.93}^{+9.24}$ & 2 & 0.0095 \\ 
HD 191939 b & $8.880324_{-0.000011}^{+0.000011}$ & $3.40_{-0.29}^{+0.31}$ & 2 & $10.00_{-0.70}^{+0.70}$ & 6 & RV & $0.89_{-0.08}^{+0.05}$ & 2 & $98.82_{-6.33}^{+4.68}$ & 2 & $6.22_{-4.65}^{+7.55}$ & 2 & 0.0930 \\ 
HD 191939 c & $28.579673_{-0.000079}^{+0.000079}$ & $2.91_{-0.30}^{+0.32}$ & 2 & $8.00_{-1.00}^{+1.00}$ & 6 & RV & $0.89_{-0.08}^{+0.05}$ & 2 & $20.80_{-1.33}^{+0.99}$ & 2 & $6.22_{-4.65}^{+7.55}$ & 2 & 0.0709 \\ 
HD 191939 d & $38.353532_{-0.000149}^{+0.000149}$ & $3.04_{-0.26}^{+0.28}$ & 2 & $2.80_{-0.60}^{+0.60}$ & 6 & RV & $0.89_{-0.08}^{+0.05}$ & 2 & $14.05_{-0.90}^{+0.67}$ & 2 & $6.22_{-4.65}^{+7.55}$ & 2 & 0.0847 \\ 
HD 207496 b & $6.440982_{-0.000015}^{+0.000015}$ & $2.33_{-0.40}^{+0.40}$ & 2 & $6.10_{-1.60}^{+1.60}$ & 7 & RV & $0.77_{-0.06}^{+0.04}$ & 2 & $77.16_{-4.47}^{+3.36}$ & 2 & $10.38_{-7.68}^{+10.52}$ & 2 & 0.0199 \\ 
... & ... &... &... &... &... &... &... &... &... &... &... &... &... \\
\hline
    \end{tabular}
    }
    \label{tab:h24_planet_masses}
\end{table}

\end{landscape}



\bsp	
\label{lastpage}
\end{document}